\documentclass{elsart}

\usepackage{graphicx}

\usepackage{amssymb}
\usepackage{amsmath}

\newcommand{\bra}[1]{\left\langle #1 \right|}
\newcommand{\ket}[1]{\left| #1 \right\rangle}
\newcommand{\erw}[3]{\bra{#1} #2 \ket{#3}}

\newcommand{\st}[1]{\left| #1 \rangle \right.}
\newcommand{\op}[2]{\left| #1 \right\rangle \left\langle #2 \right|}

\newcommand{\expec}[1]{\langle #1 \rangle}
\def\bx{{\bf x}}

\def\bE{{\bf E}}
\def\bp{{\bf p}}

\def\br{{\bf r}}

\newcommand{\eqr}[1]{Eq.~(\ref{#1})}
\newcommand{\fir}[1]{Fig.~\ref{#1}}

\begin{document}

\begin{frontmatter}

\title{The cold atom Hubbard toolbox}

\author[oxf,ibk]{D. Jaksch}, and
\author[ibk]{P. Zoller}

\address[oxf]{Clarendon Laboratory, University of Oxford, Parks Road, Oxford OX1 3PU, UK}
\address[ibk]{Institute for Theoretical Physics, University of Innsbruck, and Institute
for Quantum Optics and Quantum
Information of the Austrian Academy of Sciences, 6020 Innsbruck,
Austria}

\begin{abstract}
We review recent theoretical advances in cold atom physics
concentrating on strongly correlated cold atoms in optical
lattices. We discuss recently developed quantum optical tools for
manipulating atoms and show how they can be used to realize a wide
range of many body Hamiltonians. Then we describe connections and
differences to condensed matter physics and present applications
in the fields of quantum computing and quantum simulations.
Finally we explain how defects and atomic quantum dots can be
introduced in a controlled way in optical lattice systems.
\end{abstract}

\begin{keyword}
cold atom \sep optical lattice \sep Hubbard model \sep quantum
computing \sep atomic quantum dot

\PACS 03.75.-b \sep 32.80.Pj \sep 32.80.Qk \sep 03.67.Lx
\end{keyword}
\end{frontmatter}

\section{Introduction}
\label{Intro}

Atomic physics experiments with quantum degenerate Bose and Fermi
gases are characterized by the distinguishing features that we
have (i) a detailed microscopic understanding of the Hamiltonian
of the systems realized in the laboratory, and (ii) complete
control of the system parameters via external fields. In
particular, atoms can be trapped and their motion controlled in
magnetic and optical traps, allowing, for example, the realization
of quantum gases with different dimensionality at effectively zero
temperature. In addition, atoms have many internal states which
can be manipulated using laser light and can be employed as a
probe of the gas properties, and their collisional properties can
be tuned with magnetic and optical Feshbach resonances.

In the early days of atomic BEC experiments
\cite{NatureReview,StingReview,LeggettReview,Pitaevskii,Pethick},
the main focus was to investigate condensate properties of matter
waves like coherence, as described theoretically by the
Hartree-Fock-Bogoliubov mean field theory for weakly interacting
quantum gases. More recently, emphasis has shifted to strongly
interacting systems, which are much more in line with present
interests in theoretical condensed matter physics. In particular,
as first pointed out in Ref.~\cite{Jaksch98}, strongly interacting
systems can be realized with cold atomic gases in optical
lattices, i.e. periodic arrays of microtraps generated by standing
wave laser light fields. This leads to Hubbard type lattice
models, where atomic physics provides a whole toolbox to engineer
various types of Hamiltonians for 1D, 2D and 3D Bose and Fermi
systems which can be controlled by varying external field
parameters in a time dependent way. In addition, atomic physics
provides systematic ways of loading these lattices with atoms. A
prominent example is the Mott insulator-superfluid quantum phase
transition with cold bosonic atoms, as first observed in the
seminal experiment by I. Bloch and collaborators \cite{Bloch02}.
More generally, we expect that cold atoms in optical lattices will
be developed in the coming years as a general quantum simulator of
lattice models, allowing experimental insight into phase diagrams
for certain classes of (toy) models (such as high-$T_{c}$
superconductivity) and for parameter regimes, where no rigorous
theoretical approaches exist. In addition, new theoretical
challenges appear in this context, for example, the study of
time-dependent phenomena. Besides the condensed matter aspects,
the engineered Hubbard models have direct application in quantum
computing, where the controlled interactions can be used to create
entanglement with high fidelity.

In the present article we discuss the atomic physics point of
developing a Hubbard toolbox by controlling laser and collisional
interactions \cite{Cirac+Zoller-PhysicsToday:04}. The basic
physical mechanisms for creating periodic optical trapping
potentials are introduced in Sec.~\ref{OptLatt}. In Sec.~\ref{BHM}
we discuss some of the various Hubbard models that can be
engineered in optical lattices and also give a microscopic
explanation of the terms appearing in the Hamiltonians. Then we
point out the similarities and in particular the novel aspects in
comparison to condensed matter systems in Sec.~\ref{CMP}, and
proceed by describing possible applications in quantum computing
and quantum simulations in Sec.~\ref{QI}. Finally, some prospects
for introducing imperfections and quantum dots into optical
lattices are discussed in Sec.~\ref{AQDots} and we conclude in
Sec.~\ref{Concl}.


\section{Optical lattices}
\label{OptLatt}

Lasers are a very versatile tool for manipulating atoms. Atoms may
be cooled or can even be trapped by laser light
\cite{LaserCooling}. For trapping the purely conservative dipole
force exerted by a laser with an inhomogeneous intensity profile
is used \cite{RevJessen}. In this section we present several
examples of different laser-atom configurations which allow the
realization of a variety of trapping potential, in particular
periodic optical lattices with different geometries, and even
trapping potentials whose shape depends on the internal
(hyperfine) state of the atom. In our derivations we will neglect
spontaneous emission and later establish the consistency of this
approximation by giving an estimate for the rate at which photons
are spontaneously emitted in a typical optical lattice setup.

\subsection{Optical potentials}
\label{optpot}

The Hamiltonian of an atom of mass $m$ is given by $H_A=\bp^2/2m +
\sum_j \omega_j \op{e_j}{e_j}$. Here $\bp$ is the center of mass
momentum operator and $\st{e_j}$ denote the internal atomic states
with energies $\omega_j$ (setting $\hbar \equiv 1$). We assume the
atom to initially occupy a metastable internal state $\st{e_0}
\equiv \st{a}$ which defines the point of zero energy. The atom is
subject to a classical laser field with electric field
$\bE(\bx,t)= E(\bx,t) {\mathbf \epsilon} \exp(-i \omega t)$ where
$\omega$ is the frequency and $\mathbf \epsilon$ the polarization
vector of the laser. The amplitude of the electric field
$E(\bx,t)$ is varying slowly in time $t$ compared to $1/\omega$
and slowly in space $\bx$ compared to the size of the atom. In
this situation the interaction between atom and laser is
adequately described in dipole approximation by the Hamiltonian
$H_{\rm dip}= - {\mathbf \mu} \bE(\bx,t) + {\rm h.c.}$, where
${\mathbf \mu}$ is the dipole operator of the atom. We assume the
laser to be far detuned from any optical transition so that no
significant population is transferred from $\st{a}$ to any of the
other internal atomic states via $H_{\rm dip}$. We can thus treat
the additional atomic levels in perturbation theory and eliminate
them from the dynamics. In doing so we find the AC-Stark shift of
the internal state $\st{a}$ in the form of a conservative
potential $V(\bx)$ whose strength is determined by the atomic
dipole operator and the properties of the laser light at the
center of mass position $\bx$ of the atom. In particular $V(\bx)$
is proportional to the laser intensity $\left|E(\bx,t)\right|^2$.
Under these conditions the motion of the atom is governed by the
Hamiltonian $H=\bp^2/2m + V(\bx)$.

Let us now specialize the situation to the case where the dominant
contribution to the optical potential arises from one excited
atomic level $\st{e}$ only. In a frame rotating with the laser
frequency the Hamiltonian of the atom is approximately given by
$H_A=\bp^2/2m + \delta \op{e}{e}$ where $\delta=\omega_e - \omega$
is the detuning of the laser from the atomic transition $\st{e}
\leftrightarrow \st{a}$. The dominant contribution to the
atom-laser interaction neglecting all quickly oscillating terms
(i.e.~in the rotating wave approximation) is given by $H_{\rm
dip}=\Omega(\bx) \op{e}{a}/2 + {\rm h.c.}$. Here $\Omega=-2
E(\bx,t) \erw{e}{\mu {\mathbf \epsilon}}{a}$ is the so called Rabi
frequency driving the transitions between the two atomic levels.
For large detuning $\delta \gg \Omega$ adiabatically eliminating
the level $\st{e}$ yields the explicit expression $V(\bx)=
|\Omega(\bx)|^2/4 \delta$ for the optical potential. The
population transferred to the excited level $\st{e}$ by the laser
is given by $|\Omega(\bx)|^2 /4 \delta^2$ and this is the reason
why we require $\Omega(\bx) \ll \delta$ for our adiabatic
elimination to be valid.

\subsection{Periodic lattices}

For creating an optical lattice potential we start by
superimposing two counter propagating running wave laser beams
with $E_\pm(\bx,t)=E_0 \exp(\pm i k x)$ propagating in
$x$-direction with amplitude $E_0$, wave number $k$ and wave
length $\lambda = 2 \pi / k$. They create an optical potential
$V(\bx) \propto \cos^2(k x)$ in one dimension with periodicity
$a=\lambda/2$. Using two further pairs of laser beams propagating
in $y$ and $z$ direction respectively, a full three dimensional
periodic trapping potential of the form
\begin{equation}
\label{OptPot} V(\bx)=V_{0x} \cos^2(k x)+V_{0y} \cos^2(k y)+V_{0z}
\cos^2(k z)
\end{equation}
is realized. The depth of this lattice in each direction is
determined by the intensity of the corresponding pair of laser
beams which is easily controlled in an experiment.

{\em Bloch bands and Wannier functions:} For simplicity we only
consider one spatial dimension in \eqr{OptPot} and write down the
Bloch functions $\phi_q^{(n)}(x)$ with $q$ the quasi momentum and
$n$ the band index. The corresponding eigenenergies $E_q^{(n)}$
for different depths of the lattice $V_0/E_R$ in units of the
recoil energy $E_R=k^2/2m$ are shown in \fir{figbb}. Already for a
moderate lattice depth of a few recoil the separation between the
lowest lying bands is much larger than their extend. In this case
a good approximation for the gap between these bands is given by
the oscillation frequency $\omega_T$ of a particle trapped close
to one of the minima $x_j$ ($\equiv$ lattice site) of the optical
potential. Approximating the lattice around a minimum by a
harmonic oscillator we find $\omega_T = \sqrt{4 V_0 E_R}$
\cite{Jaksch98}.

\begin{figure}[tb]
\begin{center}
\includegraphics{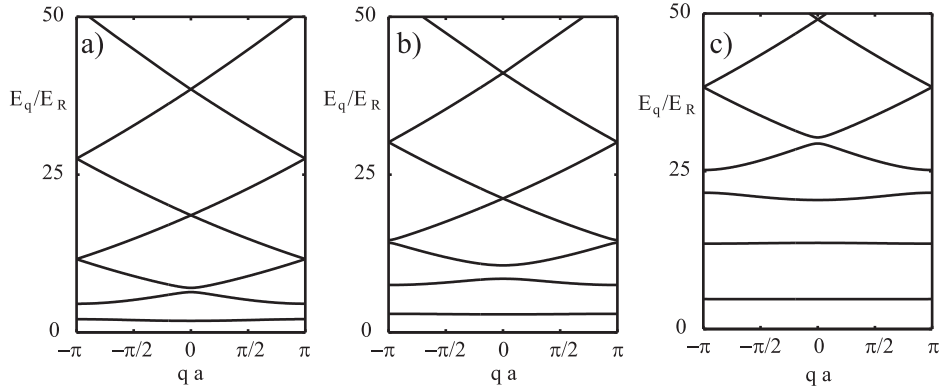}
\caption{Band structure of an optical lattice of the form
$V_0(x)=V_0 \cos^2(kx)$ for different depths of the potential. a)
$V_0=5 E_R$, b) $V_0=10 E_R$, and c) $V_0=25 E_R$.} \label{figbb}
\end{center}
\end{figure}

The dynamics of particles moving in the lowest lying well
separated bands will be described using Wannier functions. These
are complete sets of orthogonal normalized real mode functions for
each band $n$. For properly chosen phases of the $\phi_q^{(n)}(x)$
the Wannier functions optimally localized at lattice site $x_j$
are defined by \cite{Kohn}
\begin{equation}
w_{n}(x-x_j)=\Theta^{-1/2} \sum_q e^{-i q x_j} \phi_q^{(n)}(x),
\end{equation}
where $\Theta$ is a normalization constant. Note that for $V_0
\rightarrow \infty$ and fixed $k$ the Wannier function $w_{n}(x)$
tends towards the wave function of the $n$-th excited state of a
harmonic oscillator with ground state size $a_0=\sqrt{1/m
\omega_T}$. We will use the Wannier functions to describe
particles trapped in the lattice since they allow (i) to attribute
a mean position $x_j$ to the particles in a given mode and (ii) to
easily account for local interactions between particles since the
dominant contribution to the interaction energy arises from
particles occupying the same lattice site $x_j$.

\begin{figure}[tb]
\begin{center}
\includegraphics[width=13.5cm]{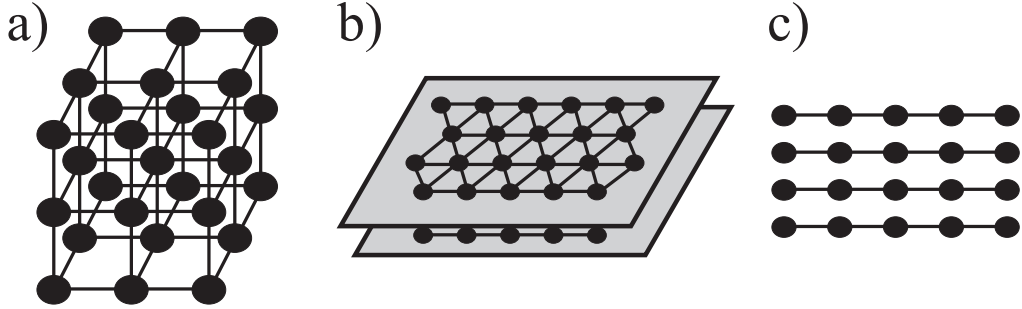}
\caption{a) Simple three dimensional cubic lattice. b) Sheets of a
two dimensional triangular lattices. c) Set of one dimensional
lattice tubes. In a)-c) tunnelling of atoms through the optical
potential barriers is only possible between sites which are
connected by lines.} \label{lattgeom}
\end{center}
\end{figure}

{\em Lattice geometry:} The lattice site positions $\bx_j$
determine the lattice geometry. For instance the above arrangement
of three pairs of orthogonal laser beams leads to a simple cubic
lattice (shown in \fir{lattgeom}a). Since the laser setup is very
versatile different lattice geometries can be achieved easily. As
one example consider three laser beams propagating at angles $2
\pi /3$ with respect to each other in the $xy$-plane and all of
them being polarized in $z$ direction. The resulting lattice
potential is given by $V(\bx) \propto 3 + 4 \cos(3 k x /2) \cos(
\sqrt 3 k y /2) + 2 \cos(\sqrt 3 k y)$ which is a triangular
lattice in two dimensions. An additional pair of lasers in $z$
direction can be used to create localized lattice sites (cf.
\fir{lattgeom}b). Furthermore, as will be discussed later in
Sec.~\ref{BHM}, the motion of the atoms can be restricted to two
or even one spatial dimension by large laser intensities. As will
be shown it is thus possible to create truly one- and
two-dimensional lattice models as indicated in \fir{lattgeom}b,c.

{\em Pseudo random site offset:} This can be achieved by
additional superlattice potentials. Two additional laser beams
propagating at small angles $\alpha$ and at $\pi - \alpha$ with
respect to the $x$- axis of the lattice, respectively, superimpose
a slowly varying potential with spatial period $l=\sin(\alpha) / 2
\bar k$ where $\bar k$ is the wave number of the additional
lasers. If $l\gg a$ the main effect of the superlattice is an
energy offset $\epsilon_i \propto \cos(x_i/l)$ for lattice site
$\bx_i$. If the periodicity $l$ is incommensurate with $a$ and
several additional lasers with different periodicity $l_n$ are
superimposed to the original lattice the site offsets $\epsilon_i$
can be made quasi random over the size of the actual optical
lattice.

{\em State dependent lattices:} As already mentioned above the
strength of the optical potential crucially depends on the atomic
dipole moment between the internal states involved. Thus we can
exploit selection rules for optical transitions to create
differing traps for different internal states of the atom
\cite{Jaksch99,Brennen99,Liu}. We will illustrate this by an
example that is particularly relevant in what follows. We consider
an atom with the fine structure shown in \fir{firlev}a, like e.g.
${}^{23}$Na or ${}^{87}$Rb, interacting with two circularly
polarized laser beams. The right circularly polarized laser
$\sigma^+$ couples the level $S_{1/2}$ with $m_s=-1/2$ to two
excited levels $P_{1/2}$ and $P_{3/2}$ with $m_s=1/2$ and
detunings of opposite sign. The respective optical potentials add
up. The strength of the resulting AC-Stark shift is shown in
\fir{firlev}b as a function of the laser frequency $\omega$. For
$\omega=\omega_L$ the two contribution cancel. The same can be
achieved for the $\sigma^-$ laser acting on the $S_{1/2}$ level
with $m_s=1/2$. Therefore at $\omega=\omega_L$ the AC--Stark
shifts of the levels $S_{1/2}$ with $m_s=\pm 1/2$ are purely due
to $\sigma_{\pm}$ polarized light which we denote by $V_\pm(x)$.
The corresponding level shifts of the hyperfine states in the
$S_{1/2}$ manifold (shown in \fir{firlev}a) are related to
$V_\pm(x)$ by the Clebsch--Gordan coefficients,
e.g.~$V_{\st{F=2,m_F=2}}(x)=V_+(x)$,
$V_{\st{F=1,m_F=1}}(x)=3V_+(x)/4  + V_-(x)/4$, and
$V_{\st{F=1,m_F=-1}}(x)= V_+(x)/4 + 3V_-(x)/4$.

\begin{figure}[tb]
\begin{center}
\includegraphics[width=13.5cm]{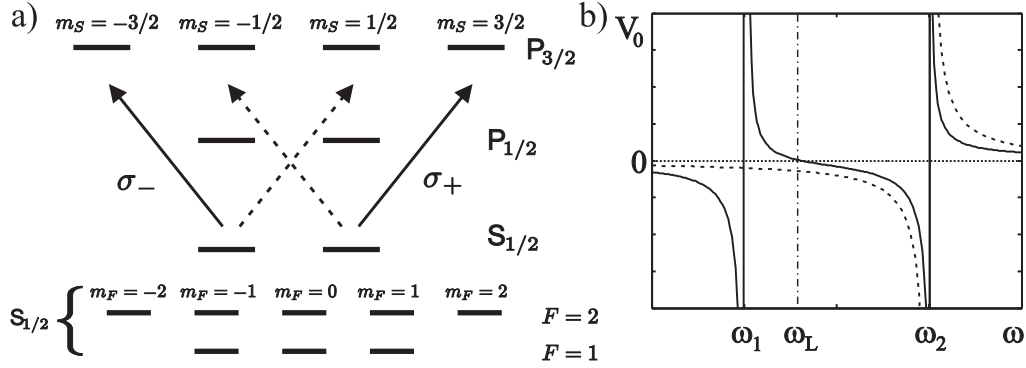}
\caption{a) Atomic fine and hyperfine structure of the most
commonly used alkali atoms $^{23}$Na and $^{87}$Rb. b) Schematic
AC--stark shift of the atomic level $S_{1/2}$ with $m_s=1/2$
(dashed curve) and with $m_s=-1/2$ (solid curve) due to the laser
beam $\sigma_+$ as a function of the laser frequency $\omega$. The
ac--stark shift of the level $S_{1/2}$ with $m_s=-1/2$ can be made
$0$ by choosing the laser frequency $\omega=\omega_L$.}
\label{firlev}
\end{center}
\end{figure}

\begin{figure}[tb]
\begin{center}
\includegraphics{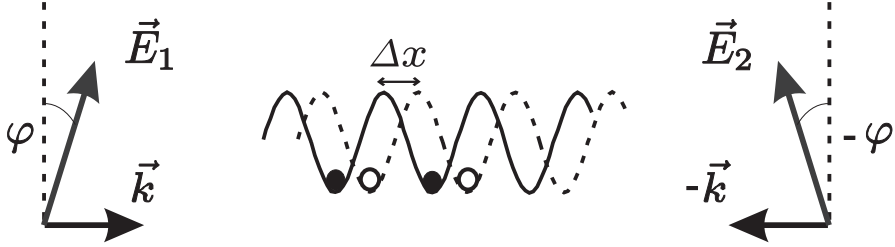}
\caption{Laser configuration for a state selective optical
potential. Two standing circular polarized standing waves are
produced out of two counter-propagating running waves with an
angle $2 \varphi$ between their polarization axes. The lattice
sites for different internal states (indicated by closed and open
circles) are shifted by $\Delta x = 2 \varphi / k$.}
\label{movesetup}
\end{center}
\end{figure}

{\em State selectively moving the lattice:} The two standing waves
$\sigma_\pm$ can be produced out of two running
counter--propagating waves with the same intensity as shown in
Fig.~\ref{movesetup}. Moreover it is possible to move nodes of the
resulting standing waves by changing the angle of the polarization
between the two running waves \cite{Jaksch99}. Let
$\{e_1,e_2,e_3\}$ be three unit vectors in space pointing along
the $\{x,y,z\}$ direction, respectively. The position dependent
part of the electric field of the two running waves ${\bf
E}_{1,2}$ is given by ${\bf E}_1 \propto e^{i k x}
\left(\cos(\varphi) e_3 + \sin(\varphi) e_2 \right)$, ${\bf E}_2
\propto e^{-i k x} \left(\cos(\varphi) e_3 - \sin(\varphi) e_2
\right)$. The sum of the two electric fields is thus ${\bf
E}_1+{\bf E}_2 \propto \cos(k x - \varphi) \sigma_- - \cos(k x +
\varphi) \sigma_+$, where $\sigma_\pm=e_2 \pm i e_3$ and the
resulting optical potentials are given by
\begin{equation}
V_\pm(x) \propto \cos^2(k x \pm \varphi).
\end{equation}
By changing the angle $\varphi$ it is therefore possible to move
the nodes of the two standing waves in opposite directions. Since
these two standing waves act as internal state dependent
potentials for the hyperfine states the optical lattice can be
moved in opposite directions for different internal hyperfine
states.

{\em Population transfer between hyperfine states:} Two atomic
hyperfine levels can be coupled coherently via magnetic dipole
moment matrix elements by an additional oscillating microwave
field. Alternatively transitions can also be driven by a Raman
laser setup which consists of two laser beams and couples two
hyperfine levels via a far detuned excited state. In both cases
the dynamics is described by a Hamiltonian of the form
$H_{R}=(\Omega_R \op{a}{b} + h.c.)/2 + \delta_R \op{b}{b}$ with
$\Omega_R$ the Rabi frequency, $\delta_R$ the detuning and
$\st{a}$, $\st{b}$ the two coupled atomic states.

\subsection{Validity}

In all of the above calculations we have only considered the
coherent interactions of an atom with laser light. Any incoherent
scattering processes which lead to spontaneous emission were
neglected. We will now establish the validity of this
approximation by estimating the mean rate $\Gamma_{\rm eff}$ of
spontaneous photon emission from an atom trapped in the lowest
vibrational state of the optical lattice. This rate of spontaneous
emission is given by the product of the life time $\Gamma$ of the
excited state and the probability of the atom occupying this
state. In the case of a blue detuned optical lattice (dark optical
lattice) $\delta < 0$ the potential minima coincide with the
points of no light intensity and we find the effective spontaneous
emission rate $\Gamma_{\rm eff} \approx -\Gamma \omega_T / 4
\delta$. If the lattice is red detuned (bright optical lattice)
$\delta >0$ the potential minima match the points of maximum light
intensity and we find $\Gamma_{\rm eff} \approx \Gamma V_0 /
\delta$. Since $V_0 > \omega_T$ the spontaneous emission in a red
detuned optical lattice will always be more significant than in a
blue lattice, however, as long as $V_0 \ll \delta$ spontaneous
emission does not play a significant role.

{\em Typical numerical values:} In a typical blue detuned optical
lattice with $\lambda=514$nm for ${}^{23}$Na atoms ($S_{1/2} -
P_{3/2}$ transition at $\lambda_2 = 589$nm and $\Gamma = 2 \pi
\times 10$MHz) the recoil energy is $E_R \approx 2 \pi  \times
33$kHz and the detuning from the atomic resonance is $\delta
\approx -2.3 \times 10^9 E_R$. A lattice with depth $V_0=25 E_R$
leads to a trapping frequency of $\omega_T = 10 E_R$ yielding a
spontaneous emission rate of $\Gamma_{\rm eff} \approx 10^{-2}/s$
while experiments are typically carried out in times shorter than
one second. Therefore spontaneous emission does not play a
significant role in such experiments.


\section{The (Bose) Hubbard model}
\label{BHM}

We consider a gas of interacting particles moving in an optical
lattice. Starting from the full many body Hamiltonian including
local two particle interactions we first give a naive derivation
of the Bose-Hubbard model (BHM) \cite{Fisher,Jaksch98} and present
related models which can be realized in an optical lattice. Then
we proceed by discussing adiabatic and irreverisble schemes for
loading the lattice with ultracold atoms. Finally we examine the
microscopic origin of the interaction terms appearing in the BHM.
Throughout we will mostly concentrate on bosonic atoms, however,
similar derivations can also be performed for fermions
\cite{Liu,Hofstetter02} and we will also give one example of such
a fermionic model.

\subsection{Naive derivation of the BHM}

The Hamiltonian of a weakly interacting gas in an optical lattice
is
\begin{equation} \label{Hfull}
H_{\rm full} =\int d^3x {\hat \Psi} ^{\dagger }({\bf x})\left(
\frac{{\bf p}^2}{2m} +V_{0}({\bf x})+V_{T}({\bf x})\right) {\hat
\Psi} ({\bf x}) +\frac{g}{2} \int dx {\hat \Psi} ^{\dagger }({\bf
x}){\hat \Psi} ^{\dagger }({\bf x}) {\hat \Psi} ({\bf x}){\hat
\Psi} ({\bf x})
\end{equation}
with ${\hat \Psi} \left({\bf x}\right)$ the bosonic field operator
for atoms in a given internal atomic state $\st{b}$ and
$V_{T}({\bf x})$ a (slowly varying compared to the optical lattice
$V_0({\bf x})$) external trapping potential, e.g.~a magnetic trap
or a superlattice potential. The parameter $g$ is the interaction
strength between two atomic particles. If the atoms interact via
$s$-wave scattering only it is given by $g=4 \pi a_s / m$ with
$a_s$ the $s$-wave scattering length. We assume all particles to
be in the lowest band of the optical lattice and expand the field
operator in terms of the Wannier functions ${\hat \Psi} ({\bf
x})=\sum_{i}{\hat b}_{i}w^{(0)}({\bf x}-{\bf x}_i)$, where ${\hat
b}_i$ is the destruction operator for a particle in site $\bx_i$.
We find $H_{\rm full}=-\sum_{i,j} J_{ij} {\hat b}_{i}^{\dagger}
{\hat b}_{j}+ \frac{1}{2} \sum_{i,j,k,l} U_{ijkl} {\hat
b}_i^\dagger {\hat b}_j^\dagger {\hat b}_k {\hat b}_l$, where
$$J_{ij}= - \int dx w_{0}({\bf x}-{\bf
x}_i)\left(\frac{p^2}{2m}+V_{0}({\bf x})+V_{T}({\bf x})\right)
w_{0}({\bf x}-{\bf x}_j),$$and $$U_{ijkl}= g \int dx w_{0}({\bf
x}-{\bf x}_i)w_{0}({\bf x}-{\bf x}_j)w_{0}({\bf x}-{\bf
x}_k)w_{0}({\bf x}-{\bf x}_l).$$

The numerical values for the offsite interaction matrix elements
$U_{ijkl}$ involving Wannier functions centered at different
lattice sites as well as tunnelling matrix elements $J_{ij}$ to
sites other than nearest neighbors (note that diagonal tunneling
is not allowed in a cubic lattice since the Wannier functions are
orthogonal) are small compared to onsite interactions $U_{0000}
\equiv U$ and nearest neighbor tunneling $J_{01} \equiv J$ for
reasonably deep lattices $V_0 \gtrsim 5 E_R$. We can therefore
neglect them and for an isotropic cubic optical lattice arrive at
the standard Bose--Hubbard Hamiltonian
\begin{equation} \label{HamBH}
H_{\rm BH}=-J \sum_{\langle i,j \rangle } {\hat b}_{i}^{\dagger}
{\hat b}_{j}+ \frac{U}{2} \sum_{j} {\hat b}_j^\dagger {\hat
b}_j^\dagger {\hat b}_j {\hat b}_j + \sum_j \epsilon_j {\hat
b}_j^\dagger {\hat b}_j.
\end{equation}
Here $\langle i,j \rangle$ denotes the sum over nearest neighbors
and the terms $\epsilon_j=V_T(\bx_j)$ arise from the additional
trapping potential. The physics described by $H_{\rm BH}$ is
schematically shown in \fir{figBHM}a. Particles gain an energy of
$J$ by hopping from one site to the next while two particles
occupying the same lattice site provide an interaction energy $U$.
An increase in the lattice depth $V_0$ leads to higher barriers
between the lattice sites decreasing the hopping energy $J$ as
shown in \fir{figBHM}b. At the same time two particles occupying
the same lattice site become more compressed which increases their
repulsive energy $U$ (cf.~\fir{figBHM}b).

\begin{figure}[tb]
\begin{center}
\includegraphics[width=13.5cm]{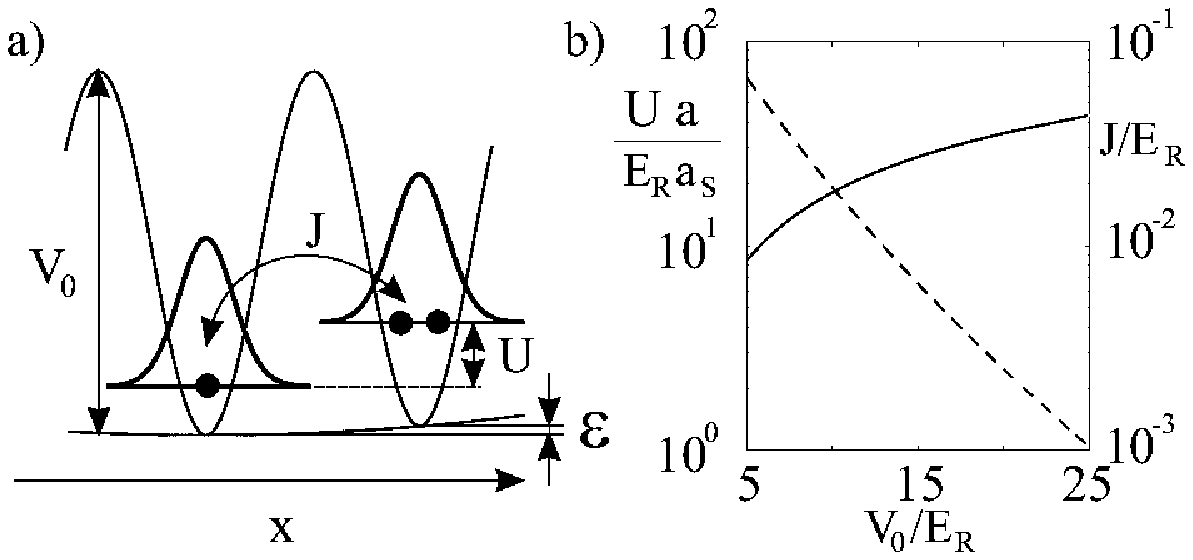}
\caption{a) Interpretation of the BHM in an optical lattice as
discussed in the text. b) Plot of scaled onsite interaction $
U/E_{R}$ multiplied by $a/a_{s}\;(\gg 1)$ (solid line with axis on
left-hand side of graph) and $J/E_{R}$ (dashed line, with axis on
right-hand side of graph) as a function of $V_{0}/E_{R}\equiv
V_{x,y,z0}/E_{R}$ (for a cubic 3D lattice). \label{figBHM}}
\end{center}
\end{figure}

{\em Tunneling term $J$:} In the case of an ideal gas where $U=0$
the eigenstates of $H_{\rm BH}$ are easily found for
$\epsilon_i=0$ and periodic boundary conditions. From the
eigenvalue equation $E_q^{(0)}=-2 J \cos(q a)$ we find that $4 J$
is the height of the lowest Bloch band. Furthermore we see that
the energy is minimized for $q=0$ and therefore particles in the
ground state are delocalized over the whole lattice, i.e. the
ground state of $N$ particles in the lattice is $\st{\Psi_{SF}}
\propto (\sum_i {\hat b}_i^\dagger)^N \st{\rm vac}$ with $\st{\rm
vac}$ the vacuum state. In this limit the system is superfluid
(SF) and possesses first order long range off diagonal
correlations \cite{Fisher,Jaksch98}.

{\em Onsite interaction $U$:} In the opposite limit where the
interaction $U$ dominates the hopping term $J$ the situation
changes completely. As discussed in detail in
\cite{Fisher,Jaksch98} a quantum phase transition takes place at
about $U \approx 5.8 z J$ where $z$ is the number of nearest
neighbors of each lattice site. The long range correlations cease
to exist in the ground state and instead the system becomes Mott
insulating (MI). For commensurate filling of one particle per
lattice site this MI state can be written as $\st{\Psi_{MI}}
\propto \prod_j {\hat b}_j^\dagger \st{\rm vac}$.

\subsection{Lattice geometry and spin models}

The lattice geometry and dimensionality are reflected in the sum
over the nearest neighbors $\langle i,j \rangle$. Exotic
geometries like e.g.~a Kagome lattice can be realized
\cite{SantosKagome} and disordered systems may be studied by
superimposing a pseudorandom potential \cite{BoseGlass}. If for
the simple cubic lattice the laser intensities in $z$ and $y$
direction are made very large compared to the intensity in $x$
tunnelling predominantly happens along the $x$ direction
(cf.~\fir{figBHM}) and the system is effectively reduced to a set
of one dimensional lattices. In addition, when the laser
intensities are turned on sufficiently slowly (adiabatically), the
correlations between atoms in different optical lattice tubes
cease via the SF to MI transition and the experimental setup
becomes an ensemble of independent one dimensional systems. These
can be used to study properties of strongly correlated gases like
e.g. a Tonks-Girardeau gas recently experimentally achieved in
\cite{ParedesTonks}, or the excitation spectrum of one dimensional
ultracold gases \cite{Stoefele}.

{\em Spin models:} If several species of atoms, e.g.~differernt
hyperfine states, are trapped in the lattice one often identifies
them with spin degrees of freedom. The most common example is a
spin $1/2$ system created from two states $\st{\downarrow} \equiv
\st{a}$ and $\st{\uparrow} \equiv \st{b}$. Such systems possess
rich phase diagrams \cite{SpinPhaseDiag} and as shown in
\cite{sorensen,Duan,Pachos} one can engineer spin exchange
interactions and design their properties like anisotropy and sign
by proper choices of optical potentials. It is even possilbe to
create Cooper pairs of bosonic atoms and realize the regime of a
Luttinger liquid \cite{Paredes}.

For a lattice of fermionic atoms, using similar setups as
discussed above and an additional Raman laser $\Omega_R$, one can
realize spin dependent Fermi Hubbard models
\cite{Hofstetter02,Liu} with e.g.~a Hamiltonian
\begin{eqnarray}
H & =&-\sum_{\sigma\langle i,j\rangle} J_{\sigma} \left(\hat
c_{\sigma i}^{\dagger } \hat c_{\sigma j}+\mathrm{h.c.}\right)
-\frac{\delta_R}{2} \sum_{i}\left(\hat c_{\uparrow
i}^{\dagger}\hat c_{\uparrow i}
- \hat c_{\downarrow i}^{\dagger} \hat c_{\downarrow i}  \right)  \nonumber \\
&& +\frac{\Omega_R}{2}\sum_{i}\left( \hat c_{\uparrow
i}^{\dagger}\hat c_{\downarrow i}+\mathrm{h.c.}\right) -U\sum_{i}
\hat c_{\uparrow i}^{\dagger} \hat c_{\downarrow i}^{\dagger} \hat
c_{\downarrow i} \hat c_{\uparrow i}, \label{eq:HubbardH:atom}
\end{eqnarray}
where $\hat c_{\sigma j}$ ($\hat c_{\sigma j}^\dagger$) destroys
(creates) a fermion in lattice site $j$ and internal state $\sigma
\in \{ \uparrow, \downarrow\}$ and the hopping matrix element
$J_{\sigma}$ can be made spin dependent by state selective optical
potentials. Also loading of mixtures of bosonic and fermionic
atoms into an optical lattice was proposed, the phase diagram was
worked out \cite{Burnett1} and effects like pairing of fermions
with bosons \cite{BoseFermiMix} or the phase separation of the two
species \cite{Buechler} were studied.

 {\em The Hofstaedter butterfly:} A slightly more involved laser
setup even allows for the realization of effective magnetic fields
in two dimensional optical lattices \cite{Hoflatt}. This requires
time reversal symmetry to be broken which can e.g.~be achieved by
a combination of accelerating a state dependent lattice and using
Raman lasers to induce the hopping between sites as schematically
shown in \fir{figHoflatt}. A particle moving around one plaquette
of the lattice acquires a phase shift that is purely determined by
the laser phases and thus can be chosen freely between $2 \pi
\alpha \in [0,2 \pi]$ \cite{Hoflatt}. It corresponds to a magnetic
flux of $\Phi = 2 \pi \alpha/e$ (with $e$ the electron charge) and
the dynamics is therefore identical to an electron moving in a
lattice subject to an external magnetic field. Any strength of the
magnetic field - an in particular also very large field strengths
- can be achieved in the optical lattice setup and thus it should
be possible to obtain the fractal energy band structure (the
Hofstaedter butterfly) predicted by Hofstaedter for lattice
electrons moving in huge magnetic fields \cite{Hoforig}. Such
optical lattice setups also allow for investigations of fractional
quantum Hall states \cite{Lukin3}.

\begin{figure}[tbp]
\begin{center}
\includegraphics[width=13.5cm]{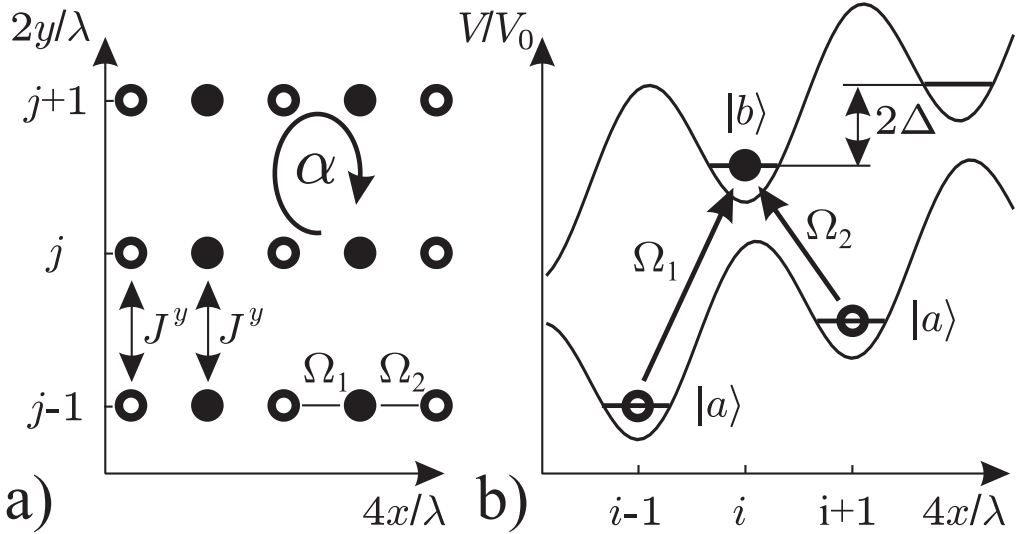}
\caption{\textit{Optical lattice setup}: Open (closed) circles
denote atoms in state $|a\rangle$ ($|b\rangle$). a) Hopping in the
$y$-direction is due to kinetic energy and described by the
hopping matrix element $J^y$ being the same for particles in
states $|a\rangle$ and $|b\rangle$. Along the $x$-direction
hopping amplitudes are due to the additional lasers $\Omega_1$ and
$\Omega_2$. b) Trapping potential in $x$-direction. Adjacent sites
are set off by an energy $\Delta$ because of the acceleration or a
static inhomogeneous electric field. The laser $\Omega_1$ is
resonant for transitions $|a\rangle \leftrightarrow |b\rangle$
while $\Omega_2$ is resonant for transitions between $|b\rangle
\leftrightarrow |a\rangle$ due to the offset of the lattice sites.
Because of a spatial dependence of $\Omega_{1,2}$ in $y$-direction
atoms hopping around one plaquette get phase shifts of $2 \pi
\alpha$. \label{figHoflatt}}
\end{center}
\end{figure}

In this article we exclusively discuss the case of tight optical
confinement in all three dimensions with few particles per lattice
site and the resulting Hubbard models. The situation where only
one or two pairs of laser beams are used to create arrays of
coupled BECs is instead similar to arrays of Josephson junctions.
Such setups have been investigated intensively both theoretically
\cite{Stoof,Smerzi,Menotti,Pitaevskii} as well as experimentally
\cite{Kasevich,Arimondo}. Also early experiments which used purely
optical cooling and loading methods to prepare atoms in the lowest
Bloch band of the lattice and hence achieved small filling factors
of approximately $10\%$ are not considered here. For such fillings
interaction effects are not important and one particle effects
like Bloch oscillations, quantum chaos, and laser cooling (see
e.g.~\cite{OldLatt1,OldLatt2}) were of main interest in these
experiments.

\subsection{Loading schemes}

Only by using atomic Bose-Einstein condensates (BEC) has it become
possible to achieve large densities corresponding to a few
particles per lattice site. A BEC can be loaded from a magnetic
trap into a lattice by slowly turning on the lasers and
superimposing the lattice potential over the trap. The system
adiabatically undergoes the transition from the SF ground state of
the BEC to the MI state for a deep optical lattice
\cite{Jaksch98}. Note that the adiabaticity condition can easily
be fulfilled in this scenario since the changing optical lattice
potential does not induce significant quasi-momenta and therefore
the relevant energy is the distance to the excited band.
Experimentally this loading scheme and the SF to MI transition was
realized in several experiments
\cite{Bloch02,Stoefele,PhillipsPattern} and led to a MI with up to
two particles per lattice site. The number of defects in the
created 'optical crystals' were limited to approximately $10\%$ of
the sites.

{\em Defect suppressed optical lattices:} One method to further
decrease the number of defects in the lattice was recently
described in \cite{Rabl03}. The proposed setup utilizes two
optical lattices for internal states $\st{a}$ and $\st{b}$ with
substantially different interaction strengths $U_b \neq U_a$ and
identical lattice site positions $\bx_i$. Initially atoms in
$\st{a}$ are adiabatically loaded into the lattice and brought
into a MI state where the number of particles per site may vary
between $n=1,...n_{\rm max}$ because of defects. Then a Raman
laser with a detuning varying slowly in time between $\delta_i$
and $\delta_f$ is used to adiabatically transfer exactly one
particle from $\st{a}$ to $\st{b}$ as shown in
\fir{defsuppfig}a,b. This is done by going through exactly one
avoided crossing. During the whole of this process transfer of
further atoms is blocked by interactions. This scheme allows a
significant suppression of defects and - with additional site
offsets $\epsilon_i$ - can also be used for patterned loading
\cite{PhillipsPattern} of the $\st{b}$ lattice.

\begin{figure}
\begin{center}
\includegraphics[width=13.5cm]{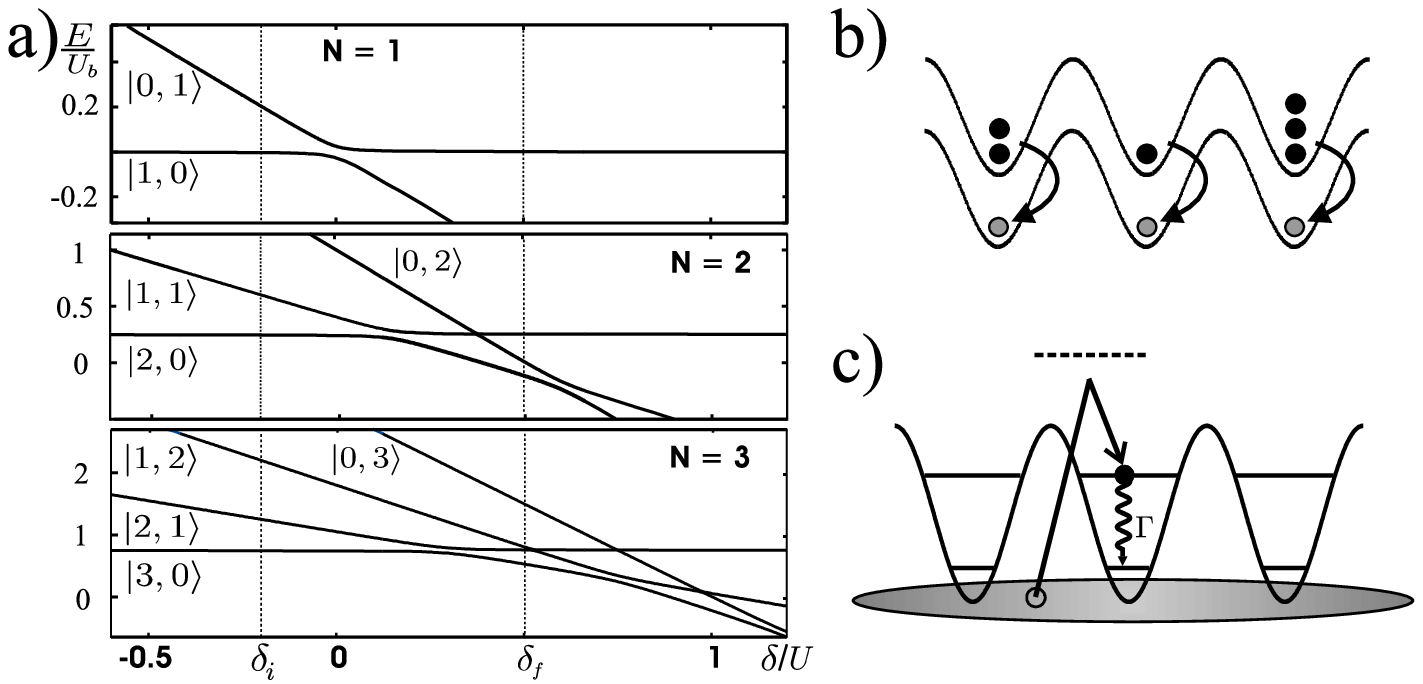}
\caption{a) Avoided crossings in the energy eigenvalues $E$ for
$n=1,2,3$ (Note the variation in the vertical scale). For the
chosen values of $\delta_i$ and $\delta_f$ (dotted vertical lines)
only one avoided crossing is traversed transferring exactly one
atom from $\st{a}$ to $\st{b}$ as schematically shown in b). c)
Loading of an atom into the first Bloch band from a degenerate gas
and spontaneous decay to the lowest band via emitting a phonon
into the surrounding degenerate gas.} \label{defsuppfig}
\end{center}
\end{figure}

{\em Irreversible loading schemes:} Further improvement could be
achieved via irreversible loading schemes. An optical lattice is
immersed in an ultracold degenerate gas from which atoms are
transferred into the first Bloch band of the lattice. By
spontaneous emission of a phonon \cite{Daley-SingAtomCoolSupe:04}
the atom then decays into the lowest Bloch band as schematically
shown in \fir{defsuppfig}. By atom-atom repulsion further atoms
are blocked from being loaded into the lattice. This scheme can
also be extended to cool atomic patterns in a lattice. In contrast
to adiabatic loading schemes it has the advantage of being
repeatable without removing atoms already stored in the lattice.

\subsection{Microscopic picture}

To obtain a better understanding of the interaction term $U$ we
consider two atoms trapped in one lattice site neglecting the
tunnelling term $J$. We approximate the lattice potential by a
harmonic oscillator which decouples the center of mass motion from
the relative motion. While the center of mass coordinate moves in
a harmonic potential of frequency $\omega_T$ the potential of the
relative coordinate $\br$ is the sum of a harmonic trap with
frequency $\omega_T$ and the interatomic interaction potential
$V_{\rm int}(\br)$ as schematically shown in \fir{micro}. The
potentials $V_{\rm int}(\br)$ are well known and one can calculate
the energy levels $\varepsilon_l$ with good precision
\cite{Julienne2,Julienne1,Weiner99}. The interaction matrix
element $U$ can then be identified with the difference in energy
$U=\varepsilon_l-3 \omega_T/2$ of the level $l$ which is closest
to the ground state of the bare harmonic oscillator level. The
states in the Bose-Hubbard picture are thus related to the
microscopic picture as indicated in \fir{micro}. Also inelastic
processes leading to particle loss can be treated in detail and a
microscopic understanding of the behavior of $U$ close to magnetic
and optical Feshbach resonances can also be gained from this
picture \cite{Julienne2,Theis,CalarcoMarker}. Based on the
detailed knowledge of the two particle levels quantum optical
schemes for the photoassociation of two atoms in $\st{a}$ to form
a molecule $\st{m}$ can be devised. For instance, as schematically
shown in \fir{micro}, using a set of Raman lasers for molecule
creation from a two particle MI state was proposed in
\cite{Jaksch02}. In comparison to customary photoassociation the
motion of the atoms and the molecule are precisely controlled and
the resulting dynamics is described by a Hamiltonian of the form
\begin{equation}
H_{\rm mol} = \left(\frac{\Omega_{m}}{2} \hat b^\dagger \hat
b^\dagger \hat m + h.c. \right) + \delta_{m} \hat m^\dagger \hat
m,
\end{equation}
with $\Omega_{\rm m}$ the effective Rabi frequency, $\delta_{\rm
m}$ the detuning and $\hat m$ ($\hat m^\dagger$) destroying
(creating) a molecule in state $\st{m}$. As above the mode
$\st{m}$ appearing in the many particle Hamiltonian $H_{\rm mol}$
has a well defined microscopic origin (cf.~\fir{micro}). In the
case of three particles occupying one lattice site analytic
understanding can be obtained from Efimov states which have been
used to calculate three particle loss rates and interaction
properties near Feshbach resonances
\cite{Petrov,Fedichev,Redthreepart}.

\begin{figure}
\begin{center}
\includegraphics[width=13.5cm]{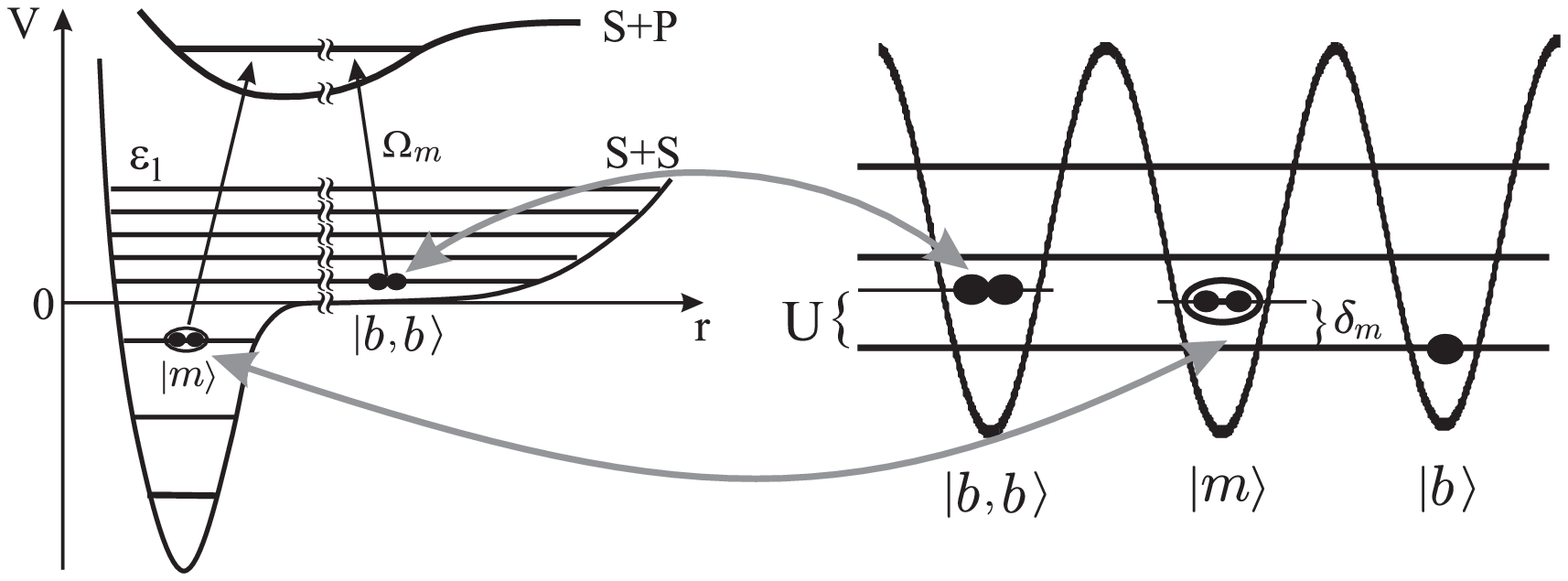}
\caption{Microscopic picture: The interatomic interaction plus
optical potential $V$ (not to scale) for the relative motion of
two atoms. Levels above $V=0$ are identified with two atoms in the
lattice and states below $V=0$ as molecular states since they
differ significantly in their extend. The relation between levels
in the Bose-Hubbard picture and in the microscopic picture are
indicated by arrows. Note that we assume the optical lattice to be
identical for atoms $\st{a}$ and the molecular state $\st{m}$
which is only true for highly excited molecular states.}
\label{micro}
\end{center}
\end{figure}


\section{Condensed matter aspects}
\label{CMP}

The setups discussed above realize - for the first time - strongly
correlated systems of neutral atoms. They are described by
Hamiltonians which have been subject to extensive investigation in
condensed matter physics where numerous techniques were developed
to find out about their properties. It is not the purpose of this
article to review these techniques but we rather point out the
most important differences between customary condensed matter
systems and atoms in optical lattices.

{\em Controllability:} The theoretical models for atoms in
lattices follow from a microscopic picture, the origin of every
term in the Hamiltonian is well justified and can be controlled
via the experimental setup. For instance, the lattice geometry and
tunnelling barriers are controlled by lasers parameters and their
arrangement. Additional flexibility comes from the ability to tune
the interaction properties of the atoms with magnetic and optical
Feshbach resonances \cite{Julienne2,Theis,CalarcoMarker}. An even
larger class of condensed matter systems can be realized by
exploiting the lattice as a universal quantum simulator as shown
in Sec.~\ref{qusim}.

{\em Inhomogeneity:} The size of an optical lattice is relatively
small in comparison to its periodicity $a$ with typically a few
hundred lattice sites in each dimension only. Also, the lattice is
usually not enclosed in a box but it is put into a slowly varying
harmonic potential which translates into a space dependent site
offset (can equally well be viewed as a local chemical potential).
This leads e.g.~to soft boundaries of the system
(cf.~\fir{figMISF}a), or the simultaneous coexistence of spatially
separated phases as shown in \fir{figMISF}b where alternating SF
and MI phases are present.

{\em Unitary dynamics:} One of the most striking differences is
the excellent isolation of an optical lattice from its
environment. As shown above for spontaneous emission, one of the
most dominant decoherence mechanisms, the typical coherence time
is on the order of seconds. The characteristic energies in the
Hamiltonian lead to a unitary evolution on millisecond time scales
and the external parameters like the laser intensity can be
changed in even shorter times. This allows the experimental study
of the unitary dynamics of strongly correlated atoms from the
adiabatic to the sudden regime \cite{Bloch02,Stephen}, a topic
usually not considered in traditional condensed matter physics. To
obtain a better understanding of these dynamical processes new
analytical and numerical methods have to be developed. Exact
numerical calculations for small systems have been performed
\cite{Jaksch02,Burnett2} and also trajectory based simulation
methods were developed \cite{NumCarosotto,Drummond}. By combining
techniques from quantum information theory with known properties
of one dimensional systems \cite{Vidal} it was possible to develop
a numerical method, the so called time evolving block decimation
algortihm (TEBD) which is related to density matrix
renormalization group (DMRG) techniques \cite{DaleyNum,Cazal}, for
simulating the unitary dynamics of a large class of one
dimensional lattice models. In \fir{figMISF}c we show results from
dynamical calculations on the SF to MI transition which indicate
differences from static ground state calculation in the
off-diagonal elements of the one particle density matrix.

{\em Quantum State engineering:} Simple condensed matter
Hamiltonians like the BHM are often considered as toy models for
obtaining insight into more complicated real physical systems and
are thus not expected to fully describe the physics and in
particular the full many body wave function. Through their exact
realization with neutral cold atoms and the excellent isolation of
these systems from their environment such toy models become
interesting for quantum state engineering and quantum computing
applications which require precise knowledge of the system wave
function. Quantum optical methods provide the accurate control and
manipulability and the strongly correlated nature of the system
can e.g.~be exploited to initialize a quantum register
\cite{Jaksch98,Jaksch99}, achieve increased stability
\cite{Dorner} or even to provide the basic resource for quantum
computations \cite{OneWay}. Some of these quantum computing
implementations will be presented in Sec.~\ref{QI}.

\begin{figure}
\begin{center}
\includegraphics[width=13.5cm]{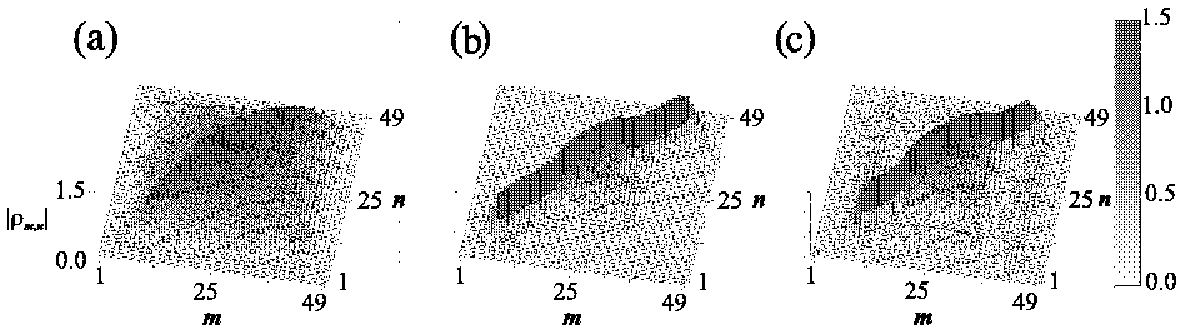}
\caption{Absolute values of the one particle density matrix
$\rho_{m,n}=\expec{\hat b_m^\dagger \hat b_n}$ as a function of
site indices $m,n$ for a one dimensional lattice superimposed by
an additional slowly varying harmonic oscillator potential. All
results are obtained by the TEBD \cite{Stephen}. (a) SF ground
state for $U/2J = 2$, (b) Intermediate ground state for $U/2J=6$.
This state clearly indicates the coexistence of MI and SF states
in different regions of the system. The SF region in the center of
the trap is surrounded by two MI regions with one particle per
lattice site followed by a small superfluid at the edges of the
trap. (c) Intermediate state created from the SF by dynamically
ramping up the depth of the lattice until $U/2J=6$ \cite{Stephen}.
Note the correlations between the two SF regions which are absent
in the corresponding ground state.} \label{figMISF}
\end{center}
\end{figure}


\section{Quantum computing implementations}
\label{QI}

One of the big challenges for 21st century physics is the
demonstration of a quantum computer which promises a significant
speed up for certain kinds of algorithms, most prominently the
factorization of large numbers and searching in unsorted databases
\cite{Nielsen+Chuang-QC:00}. While the realization of a universal
quantum computer is certainly a long term goal, also a number of
{\em nontrivial} applications with limited quantum computer
resources exist for the nearer future in the areas of quantum
communication, entanglement creation (see Sec.~\ref{QuColl}),
special purpose quantum computing, or Feynman's universal quantum
simulator (see Sec.~\ref{qusim}).

The basic requirements for a physical system to be capable of
quantum computing have been laid out by DiVincenzo
\cite{DiVincenzo00}. Many of these are fulfilled for atoms in
optical lattices. An optical lattice provides an intrinsically
{\em scalable} and {\em well defined} set of qubits (quantum
register) with {\em long decoherence} times. These are usually
constituted by two metastable ground states of the atoms $\st{a}$
and $\st{b}$ trapped in the lattice. The {\em initialization} of
the quantum register can be achieved by the loading schemes
discussed above where the MI state with one particle per lattice
site corresponds to the register initialized with zeros. Any
quantum computation can be carried out by applying {\em single
qubit gates} via Raman lasers, and controlled interactions between
the particles to perform two qubit gates as described in
Sec.~\ref{QuColl} \cite{Jaksch99,Brennen99,Brennen00,Bloch03}.
Also several other schemes for quantum computing in optical
lattices have been proposed based on e.g.~strong dipole-dipole
interactions between Rydberg atoms \cite{Jaksch00}, the motional
state of the atoms \cite{LewenMotion}, atom tunnelling between
neighboring sites \cite{Pachos1}, or measurements on highly
entangled states \cite{OneWay}. Currently the main (technical)
drawback in many of these optical lattice quantum computing
schemes is the difficulty of addressing individual atoms. While a
large number of quantum gate operations can already be carried out
in parallel \cite{Bloch03} it is not yet experimentally possible
to perform them on selected pairs of atoms. However, proposals on
utilizing marker atoms \cite{CalarcoMarker} or on quantum
computing via global system control \cite{Benjamin} exist and will
allow to overcome this problem in the near future.

\subsection{Entanglement creation via coherent ground state collisions}
\label{QuColl}

 We exploit the two particle interaction term $g \int d^3x
{\hat \Psi} ^{\dagger }({\bf x}){\hat \Psi} ^{\dagger }({\bf x})
{\hat \Psi} ({\bf x}){\hat \Psi} ({\bf x})/2$ from \eqr{Hfull}
whose origin lies in the atom-atom interaction potential for the
controlled creation of entangled states. In optical lattices these
nonlinear atom-atom interactions can be large \cite{Jaksch99},
even for interactions between individual pairs of atoms. In
addition we use a state selective optical lattice potential as
introduced in Sec.~\ref{OptLatt} to provide control over the
motional states of the atoms.

Consider a situation where two atoms populating the internal
states $|a\rangle $ and $|b\rangle $, respectively, are trapped in
the ground states $\psi _{0}^{a,b}$ of two potential wells
$V^{a,b}$. Initially, at time $t=-\tau$ , these wells are centered
at positions $\bar{x}^{a}$ and $\bar{x}^{b}$, sufficiently far
apart (distance $d=\bar{x}_{b}-\bar{x}_{a}$) so that the particles
do not interact (see \fir{pzfigatom1}a). The positions of the
potentials are moved along trajectories $\bar{x}^{a}(t)$ and
$\bar{x}^{b}(t)$ so that the wave packets of the atoms overlap for
certain time, until finally they are restored to the initial
position at $t=\tau$. This situation is described by the
Hamiltonian
\begin{equation}
H=\sum_{\beta =a,b}\left[ \frac{({p}^{\beta })^{2}}{2m}+V^{\beta
}\left( {x}^{\beta }\!-\!\bar{x}^{\beta }(t)\right) \right]
+V_{\rm int}^{\mathrm{ ab}}({x}^{a}\!-\!{x}^{b}). \label{Hamil}
\end{equation}
Here, $V^{a,b}\left( \hat{x}^{a,b}-\bar{x}^{a,b}(t)\right)$
describe the displaced trap potentials and $V_{\rm int}^{\mathrm{
ab}}$ is the interatomic interaction potential. Ideally, we want
to implement the transformation from before to after the
collision,
\begin{equation}
\psi _{0}^{a}(x^{a}\!-\!\bar{x}^{a})\psi _{0}^{b}(x^{b}\!-\!\bar{x}%
^{b})\rightarrow e^{i\phi }\psi
_{0}^{a}(x^{a}\!-\!\bar{x}^{a})\psi
_{0}^{b}(x^{b}\!-\!\bar{x}^{b}),  \label{transf}
\end{equation}
where each atom remains in the ground state of its trapping
potential and preserves its internal state. The phase $\phi
=\phi^{a}+\phi^{b}+\phi^{\mathrm{ab}}$ will contain a contribution
$\phi^{\mathrm{ab}}$ from the interaction (collision) and
(trivial) single particle kinematic phases $\phi^{a}$ and
$\phi^{b}$. The transformation (\ref{transf}) can be realized in
the \emph{adiabatic limit}, whereby we move the potentials slowly
on the scale given by the trap frequency, so that the atoms remain
in their ground state. In this case the collisional phase shift is
given by $\phi ^{\mathrm{ab}}=\int_{-\infty }^{\infty }dt\Delta
E(t)$, where $\Delta E(t)$ is the energy shift induced by the
atom--atom interactions, which in the case of purely $s$-wave
scattering is given by
\begin{equation}
\Delta E(t)=\frac{4\pi a_{s}}{m}\int dx|\psi _{0}^{a}\left(
x-\bar{x}^{a}(t)\right) |^{2}|\psi _{0}^{b}\left(
x-\bar{x}^{b}(t)\right) |^{2}. \label{deltaE}
\end{equation}
In addition we assume that $|\Delta E(t)|\ll \omega_T $ so that no
sloshing motion is excited.

\begin{figure}[tbp]
\begin{center}
\includegraphics[width=8.0cm]{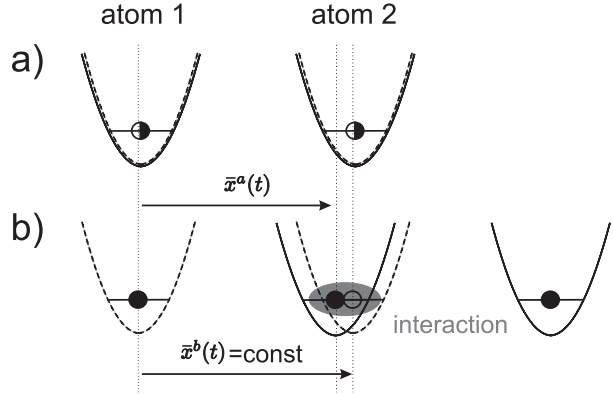}
\caption{We collide one atom in internal state $\st{a}$ (filled
circle, potential indicated by solid curve) with a second atom in
state $\st{b}$ (open circle, potential indicated by dashed
curve)). In the collision the wave function accumulates a phase
according to Eq.~(\protect\ref{transf}). a) Configurations at
times $t=\pm \tau$ and b) at time $t$.} \label{pzfigatom1}
\end{center}
\end{figure}

This process of {\em colliding atoms by hand} can be used to
create entangled states between the two atoms. By moving a state
selective optical lattice potential the wave function of each
atomic qubit splits up in space according to the internal
superposition of states $\st{a}$ and $\st{b}$ as shown in cf.
\fir{pzfigatom1}b. Only the wave function of the left atom in
state $\st{a}$ reaches the position of the second atom in state
$\st{b}$ and they will interact with each other. However, any
other combination of internal states will not interact and
therefore such a collision in a state selective optical lattice is
conditional on the internal state.

A maximally entangled Bell state can be created by carrying out a
state selective collision with a phase $\phi ^{\rm{ab}}=\pi$ and
an initial superposition $\st{+} \propto \st{a}+\st{b}$ of each
atom. After the collision this state is transformed into
$\st{a}\st{+} + \st{b}\st{-}$ with $\st{-} \propto \st{a}
-\st{b}$, i.e.~$\st{+}$ of the second atom is flipped to the
orthogonal state $\st{-}$ conditional on the state of the first
atom thus creating a Bell state. Identifying $\st{a} \equiv
\st{0}$ and $\st{b} \equiv \st{1}$ we obtain for the logical
states the mapping $\st{i,j} \rightarrow (-1)^{(1-i)j} \st{i,j}$,
with $i,j \in \{0,1\}$ (up to trivial phases) which realizes a
two-qubit phase gate. This gate in combination with single-qubit
rotations is universal i.e.~they allow to implement any unitary
dynamics on a quantum register \cite{Nielsen+Chuang-QC:00}.


\subsection{Universal Quantum Simulators}

\label{qusim}

Building a general purpose quantum computer, which is able to run
for example Shor's algorithm, requires quantum resources which
will only be available in the long term future. Thus, it is
important to identify \emph{nontrivial} applications for quantum
computers with limited resources, which are available in the lab
at present. Such an example is provided by \emph{Feynman's
universal quantum simulator} (UQS)
\cite{Lloyd-UniversalQuantumSimulators:96}. A UQS is a controlled
device that, operating itself on the quantum level, efficiently
reproduces the dynamics of any other many-particle system that
evolves according to short range interactions. Consequently, a UQS
could be used to efficiently simulate the dynamics of a generic
many-body system, and in this way function as a fundamental tool
for research in many body physics, e.g.~to simulate spin systems.

According to Jane \emph{et al.} \cite{JaneETAL:03} the very nature
of the Hamiltonian available in quantum optical systems makes them
best suited for simulating the evolution of systems whose building
blocks are also two-level atoms, and having a
Hamiltonian$H_{N}=\sum_{a}H^{(a)}+\sum_{a\neq b}H^{(ab)}$ that
decomposes into one-qubit terms $H^{(a)}$ and two-qubit terms
$H^{(ab)}$. A starting observation concerning the simulation of
quantum dynamics is that if a Hamiltonian $K=\sum_{j=1}^{s}K_{j}$
decomposes into terms $K_{j}$ acting
in a small constant subspace, then by the Trotter formula $e^{-iK\tau}%
=\lim_{m\rightarrow\infty}\left(
e^{-iK_{1}\tau/m}e^{-iK_{2}\tau/m}\ldots e^{-iK_{s}\tau/m}\right)
^{m}$ we can approximate an evolution according to $K$ by a series
of short evolutions according to the pieces $K_{j}$. Therefore, we
can simulate the evolution of an $N$-qubit system according to the
Hamiltonian $H_{N}$ by composing short one-qubit and two-qubit
evolutions generated, respectively, by $H^{(a)}$ and $H^{(ab)}$.
In quantum optics an evolution according to one-qubit Hamiltonians
$H^{(a)}$ can be obtained directly by properly shining a laser
beam on the atoms or ions that host the qubits. Instead, two-qubit
Hamiltonians are achieved by processing some given interaction
$H_{0}^{(ab)}$ (see the example below) that is externally enforced
in the following way. Let us consider two of the $N$ qubits, that
we denote by $a$ and $b$. By alternating evolutions according to
some available, switchable two qubit interaction $H_{0}^{(ab)}$
for some time with local unitary
transformations, one can achieve an evolution%
\[
U(t=\sum_{j=1}^{n}t_{j})=\prod_{j=1}^{n}V_{j}\exp(-iH_{0}^{(ab)}t_{j}%
)V_{j}^{\dag}=\prod_{j=1}^{n}\exp(-iV_{j}H_{0}^{(ab)}V_{j}^{\dag}t_{j})
\]
where $t=\sum_{j=1}^{n}t_{j}$, $V_{j}=u_{j}^{(a)}\otimes
v_{j}^{(b)}$ with $u_{j}$ and $v_{j}$ being one-qubit unitaries.
For a small time interval
$U(t)\simeq1-it\sum_{j=1}^{n}p_{j}V_{j}H_{0}^{(ab)}V_{j}^{\dag}+O(t^{2})$
with $p_{j}=t_{j}/t$, so that by concatenating several short gates
$U(t)$,
$U(t)=\exp(-iH_{\text{eff}}^{(ab)}t)+O(t^{2}),$we can simulate the Hamiltonian%
\[
H_{\text{eff}}^{(ab)}=\sum_{j=1}^{n}p_{j}V_{j}H_{0}^{(ab)}V_{j}^{\dag}+O(t)
\]
for larger times. Note that the systems can be classified
according to the availability of homogeneous manipulation,
$u_{j}=v_{j}$, or the availability of local individual addressing
of the qubits, $u_{j}\neq v_{j}$.

Cold atoms in optical lattices provide an example where single
atoms can be loaded with high fidelity into each lattice site, and
where cold controlled collisions provide a way of entangling these
atoms in a highly parallel way. This assumes that atoms have two
internal (ground) states $|0\rangle \equiv|\downarrow\rangle$ and
$|1\rangle\equiv|\uparrow\rangle$ representing a qubit, and that
we have two \textit{spin-dependent} lattices, one trapping the
$|0\rangle$ state, and the second supporting the $|1\rangle$. An
interaction between adjacent qubits is achieved by displacing one
of the lattices with respect to the other as discussed in the
previous subsection. In this way the $|0\rangle$ component of the
atom $a$ approaches in space the $|1\rangle$ component of atom
$a+1$, and these collide in a controlled way. Then the two
components of each atom are brought back together. This provides
an example of implementing an Ising $\sum_{a\neq
b}H_{0}^{(ab)}=\sum_{a}\sigma_{z}^{(a)}\otimes\sigma _{z}^{(a+1)}$
interaction between the qubits, where the $\sigma^{(a)\prime}$s
denote Pauli matrices. By a sufficiently large, relative
displacement of the two lattices, also interactions between more
distant qubits could be achieved. A local unitary transformation
can be enforced by shining a laser on the atoms, inducing an
arbitrary rotation between $|0\rangle$ and $|1\rangle$. On the
time scale of the collisions requiring a displacement of the
lattice (the entanglement operation) these local operations can be
assumed instantaneous. In the optical lattice example it is
difficult to achieve an individual addressing of the qubits. Such
an addressing would be available in an ion trap array as discussed
in Ref.~\cite{Cirac+Zoller-Frontiers:04}. These operations provide
us with the building blocks to obtain an effective Hamiltonian
evolution by time averaging as outlined above.

As an example let us consider the ferromagnetic
[antiferromagnetic] Heisenberg
Hamiltonian%
$
H=J\sum_{j=x,y,z}\sigma_{j}\otimes\sigma_{j}%
$ where $J>0$ [$J<0$]. An evolution can be simulated by short
gates with $H_{0}^{(ab)}=\gamma\sigma_{z}\otimes\sigma_{z}$
alternated with local unitary
operations%
\begin{align*}
p_{1}  &  =\frac{1}{3},\quad V_{1}=\hat{1}\otimes\hat{1}\\
p_{2}  &  =\frac{1}{3},\quad V_{2}=\frac{\hat{1}-i\sigma_{x}}{\sqrt{2}}%
\otimes\frac{\hat{1}-i\sigma_{x}}{\sqrt{2}}\\
p_{3}  &  =\frac{1}{3},\quad V_{3}=\frac{\hat{1}-i\sigma_{y}}{\sqrt{2}}%
\otimes\frac{\hat{1}-i\sigma_{y}}{\sqrt{2}}%
\end{align*}
without local addressing, as provided by the standard optical
lattice setup. The ability to perform independent operations on
each of the qubits would translate into the possibility to
simulate any bipartite Hamiltonians.

\begin{figure}[ptb]
\begin{center}
\includegraphics[width=13.5cm]{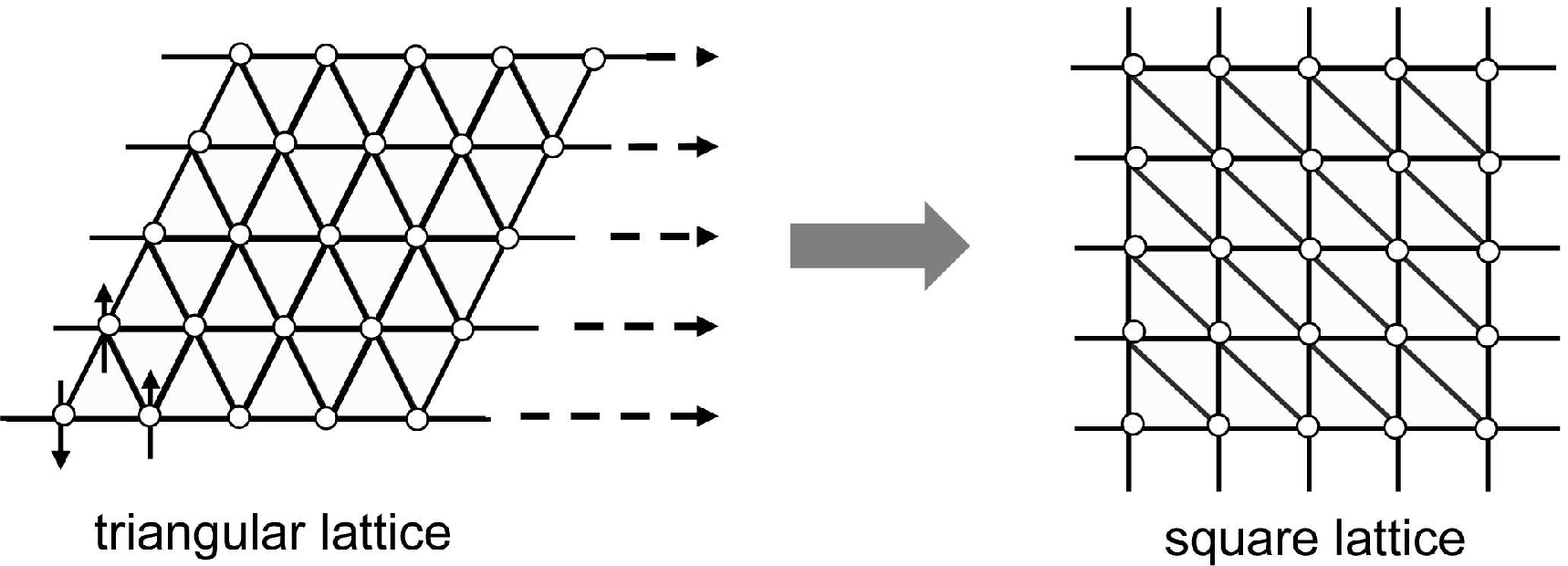}
\caption{Illustration how triangular configurations of atoms with
nearest neighbor interactions may be simulated in a rectangular
lattice. \label{fig2DLattice}}
\end{center}
\end{figure}

An interesting aspect is the possibility to simulate effectively
different lattice configurations: for example, in a 2D pattern a
system with nearest neighbor interactions in a triangular
configuration can be obtained from a rectangular array
configuration. This is achieved by making the subsystems in the
rectangular array interact not only with their nearest neighbor
but also with two of their next-to-nearest neighbors in the same
diagonal (see Fig.~\ref{fig2DLattice}).

One of the first and most interesting applications of quantum
simulations is the study of quantum phase transitions
\cite{Sachdev-QuantPhaseTransition:99}. In this case one would
obtain the ground state of a system, adiabatically connecting
ground states of systems in different regimes of coupling
parameters, allowing to determine its properties.

\section{Impurities and Atomic Quantum Dots}

\label{AQDots}

\begin{figure}[ptb]
\begin{center}
\includegraphics[width=13.5cm]{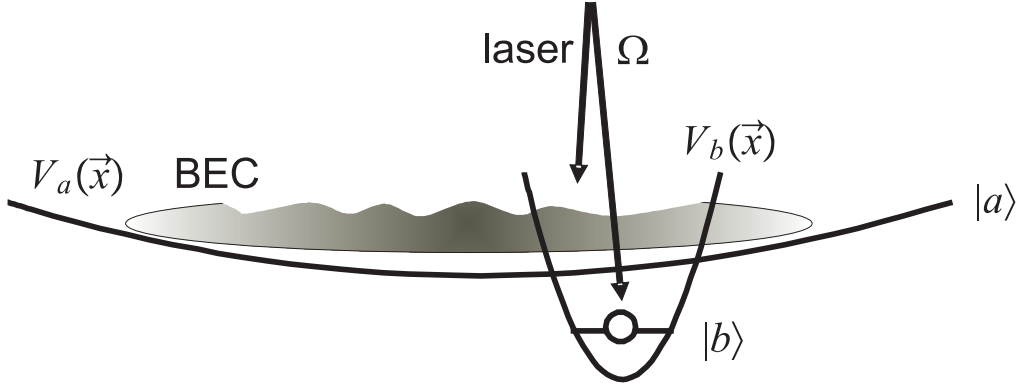}
\caption{Schematic setup of an atomic quantum dot coupled to a
superfluid atomic reservoir. The Bose-liquid of atoms in state
$\st{a}$ is confined in a shallow trap $V_{a}(\mathbf{x})$. The
atom in state $\st{b}$ is localized in tightly confining potential
$V_{b}(\mathbf{x})$. Atoms in state $\st{a}$ and $\st{b}$ are
coupled via a Raman transition with effective Rabi frequency
$\Omega$. A large onsite interaction $U_{bb}>0$ allows only a
single atom in the dot.} \label{cap:setup}
\end{center}
\end{figure}

A focused laser beam superimposed to a trap holding a quantum
degenerate bose or fermi gas can form a tight local trapping
potential, which can play the role of an \emph{impurity site}, or
an \emph{atomic quantum dot}, coupled to a cold atom reservoir
\cite{Diener+RaizenETAL-AQD:02,RecatiETAL:04}. This opens the
possibility of studying impurity physics or the analogue of
quantum dot physics with quantum degenerate gases with
controllable external parameters, and has several interesting
applications, e.g. extraction of single atoms from an atomic BEC.
Impurity potentials can be created as a local deformation of the
original trap holding the atomic gas with tunnel coupling between
reservoir and the dimple, or with a spin-dependent potential. In
the latter case, atoms in the internal state $|a\rangle$ see only
the large trap representing the reservoir, while atoms in state
$|b\rangle$ are held by the local tight trap. Coupling of atoms
occupying the atomic quantum dot to the reservoir of $b$-atoms is
achieved via a Raman transition (Fig.~\ref{cap:setup}). Depending
on the choice of the trapping potential for the reservoir atoms,
we can realize configurations, where the dot is coupled to a
reservoir of a 1D (Luttinger), 2D or 3D quantum gas. Atoms
occupying the dot and the reservoir will interact via collisional
interactions. In particular, in the limit of a tight trap
representing the dot and strong repulsion between the atoms we can
have a situation of a \emph{collisional blockade regime}
\cite{RecatiETAL:04}, where only a single atom can occupy the dot,
in analogy to the Coulomb blockade in electronic quantum dots. In
the following we will consider two specific examples to illustrate
these configurations. Again our emphasis will be on the derivation
of the corresponding Hamiltonians, while we refer to the original
papers for a discussion of the system dynamics.

\subsection{Atomic quantum dot coupled to superfluid reservoir: quantum
engineering of a spin-boson model}

As our first example, we consider the configuration outlined in
Fig.~\ref{cap:setup} where a superfluid reservoir of Bose atoms in
internal state $\st{a}$ is coupled via Raman transitions to the
lowest vibrational state of an atomic quantum dot with atoms in
state $\st{b}$ \cite{RecatiETAL:04}. The resulting dynamics will
be described by a \emph{spin-Boson model} \cite{LeggettETAL:87},
where the occupation of the dot (represented by a pseudo-spin) is
coupled to the density and phase fluctuations of the superfluid.
The distinguishing features of the model are: (i) the control of
the spin-phonon bath coupling by the collisional parameters, and
(ii) the realization of an ohmic or superohmic reservoir. As a
result, the rich dynamics for the spin-Boson model, for example
dissipative phase transitions, should be observable in the quantum
dot dynamics.

The collisional interaction of atoms in the two internal levels\
are described by a set of parameters $g_{\alpha\beta}=4\pi
a_{\alpha\beta}/m$ with scattering lengths $a_{\alpha\beta}$
($\alpha,\beta=a,b$). Following steps analogous to the derivation
of the BHM Hamiltonian in Sec.~\ref{BHM} we arrive at the
Hamiltonian for the dot $H_{b}$ and its coupling to the reservoir
$H_{ab}$,
\begin{align}
\label{eq:bHam}H_{b}+H_{ab}  &  =\left(  -\delta+g_{ab}\int
d\mathbf{x~|}\psi_{b}(\mathbf{x})|^{2}\hat{\rho}_{a}(\mathbf{x})\right)
\hat{b}^{\dagger}\hat{b}\nonumber\\
&
+\frac{U_{bb}}{2}\hat{b}^{\dagger}\hat{b}^{\dagger}\hat{b}\hat{b}+\int
d\mathbf{x}\Omega(\hat{\Psi}_{a}(\mathbf{x})\psi_{b}(\mathbf{x})\hat
{b}^{\dagger}+\mathrm{h.c.}),
\end{align}
Here $\hat{b}$ is the bosonic destruction operator for $b$-atom in
the lowest vibrational state of the dot with wave function
$\psi_{b}(\mathbf{x})$, while $\hat{\Psi}_{a}(\mathbf{x})$ is the
annihilation operator for a reservoir atom $a$ at point
$\mathbf{x}$. The first term in (\ref{eq:bHam}) contains the Raman
detuning $\delta$ and the collisional shift of the lowest
vibrational state of the dot due to interaction with the reservoir
with atomic density
$\hat{\rho}_{a}(\mathbf{x})=\hat{\Psi}_{a}^{\dagger}(\mathbf{x})\hat{\Psi}%
_{a}(\mathbf{x})$. The second term is the onsite repulsion
$U_{bb}$ of the atoms occupying the quantum dot, and the third
term is the laser induced coupling transferring atoms from the dot
to the reservoir with effective two-photon Rabi frequency
$\Omega$.

At sufficiently low temperatures the reservoir atoms in $\st{a}$
form a superfluid Bose liquid with an equilibrium liquid density
$\rho_{a}$. The only available excitations at low energies are
then phonons with linear dispersion $\omega=v_{s}|\mathbf{q}|$ and
sound velocity $v_{s}$. We write the atomic density operator in
terms of the density fluctuation operator $\hat{\Pi}$,
$\hat{\rho}_{a}(\mathbf{x})=\rho_{a}+\hat{\Pi}(\mathbf{x})$, which
is canonically conjugate to the superfluid phase
$\hat{\phi}(\mathbf{x})$ with $\hat{\Psi}_{a}$
$\sim\hat{\rho}_{a}{}^{1/2}e^{-i\hat{\phi}}$. In the long
wavelength approximation the dynamics of the superfluid is
described by a (quantum) hydrodynamic Hamiltonian, $H_{a}=
v_{s}\sum_{\mathbf{q}}|\mathbf{q}|\hat b_{\mathbf{q}}^{\dagger}
\hat b_{\mathbf{q}}$, which has the form of a collection of
harmonic sound modes with annihilation operators of phonons $\hat
b_{\mathbf{q}}$. Since the excitations of a weakly interacting
Bose-liquid are phonon-like only for wavelengths larger than the
healing length $\xi\geq l_{b}$, the summation over the phonon
modes is cutoff at a frequency $\omega_{c}=v_{s}/\xi\approx
g_{aa}\rho_{a}$.

In the collisional blockade limit of large onsite interaction
$U_{bb}$ only states with occupation $n_{b}=0$ and $1$ in the dot
participate in the dynamics, while higher occupations are
suppressed by the large collisional shift. This situation and its
description is analogous to the Mott insulator limit in optical
lattices. Thus, the quantum state of a dot is described by a
pseudo-spin-$1/2$, with the spin-up or spin-down state
corresponding to occupation by a single or no atom in the dot.
Using standard Pauli matrix notation, the dot occupation operator
$\hat{b}^{\dagger}\hat{b}$ is then replaced by $(1+\sigma_{z})/2$
while $\hat{b}^{\dagger}\rightarrow\sigma_{+}$. Furthermore, for
the long wavelength wavevectors $|\mathbf{q}|l_{b}\ll1$ with
$l_{b}$ the dimension of the dot, the phonon field operators may
be replaced by their values at $\mathbf{x}=\mathbf{0}$. This gives
the Hamiltonian
\begin{equation}
H_{b}+H_{ab}=\left( -\frac{\delta}{2}+\frac{g_{ab}}{2}\hat{\Pi
}(\mathbf{0})\right)  \sigma_{z}+\frac{\Delta}{2}(\sigma_{+}%
e^{-i\hat{\phi}(\mathbf{0})}+\mathrm{h.c.}). \label{eq:Hbnew}%
\end{equation}
Here $\Delta\sim\Omega n_{a}^{1/2}$ is an effective Rabi frequency
with $n_{a}$ the number of bosons inside the quantum dot
wavefunction $\psi_{b}$. An important feature of this Hamiltonian
is\ that phonons couple to the dot dynamics via (i) the
collisional interactions coupling to the density fluctuations
$\hat{\Pi}(\mathbf{0})$, and (ii) to the superfluid phase
$\hat{\phi}(\mathbf{0})$ via the laser couplings. A unitary
transformation
$H=S^{-1}(H_{a}+H_{b}+H_{ab})S$ with $S=\exp(-\sigma_{z}i\hat{\phi}%
(\mathbf{0})/2)$ transforms the phase fluctuations in the Rabi
term to a frequency fluctuation which adds this contribution to
the first term in (\ref{eq:Hbnew}). Thus the dynamics of the
quantum dot coupled to the superfluid reservoir gives rise to a
\emph{spin-Boson Hamiltonian} \cite{LeggettETAL:87}
\begin{equation}
H=\sum_{\mathbf{q}}\omega_{\mathbf{q}}b_{\mathbf{q}}^{\dagger
}b_{\mathbf{q}}+\left(  -\delta+\sum_{\mathbf{q}}\lambda_{\mathbf{q}%
}(b_{\mathbf{q}}+b_{\mathbf{q}}^{\dagger})\right)  \frac{\sigma_{z}}%
{2}-\frac{\Delta}{2}\sigma_{x} \label{eq:Hspinboson}%
\end{equation}
with amplitudes of the total phonon coupling
$\lambda_{\mathbf{q}}=\left\vert m
\mathbf{q}v_{s}^{3}/2V\rho_{a}\right\vert ^{1/2}\left( g_{ab}\rho
_{a}/mv_{s}^{2}-1\right)  $, as a sum of the collisional
interaction $g_{ab}$ and the transformed phase fluctuations. We
note the following two features of the model.

\emph{Control of spin-bath coupling by quantum interference:} For
a repulsive inter-species interaction, $g_{ab}>0$, both
contributions in $\lambda_q$ interfere destructively: the effect
of the phonon excitation in a laser-driven transition
$a\leftrightarrow b$ can be precisely cancelled by the change in
the direct \textquotedblleft elastic\textquotedblright\
interaction between the liquid and the $b$ atoms, as described by
the first term in Eq.~(\ref{eq:Hbnew}). In a weakly interacting
gas $mv_{s}^{2}=\rho_{a}g_{aa}$ and thus the coupling constants
$\lambda_{\mathbf{q}}$ vanish at $g_{ab}=g_{aa}$. At this special
point, we have thus decoupled the dissipative phonon reservoir
from the spin, i.e. we have stable superposition of occupation and
non-occupation of the dot, exhibiting undamped Rabi oscillations
of the quantum dot's occupancy.

\emph{Ohmic and superohmic bath couplings:} The system is
characterized by the
effective density of states $J(\omega)=\sum_{\mathbf{q}}\lambda_{\mathbf{q}%
}^{2}\delta(\omega-\omega_{\mathbf{q}})=2\alpha\omega^{s}$ where
$\alpha \sim(g_{ab}\rho_{a}/mv_{s}^{s}-1)^{2}$ is the dissipation
strength due to the spin-phonon coupling and $D=s$ the dimension
of the superfluid reservoir. In the standard terminology
\cite{LeggettETAL:87}, $s=1$ and $s>1$ correspond to the ohmic and
superohmic cases, respectively.

For a discussion of the dynamics and the validity of the model,
and specific atomic realizations we refer to \cite{RecatiETAL:04}
and \cite{LeggettETAL:87}.

\subsection{A ``Single Atom Transistor'' with a 1D optical lattice}

\begin{figure}[ptb]
\begin{center}
\includegraphics[width=13.5cm]{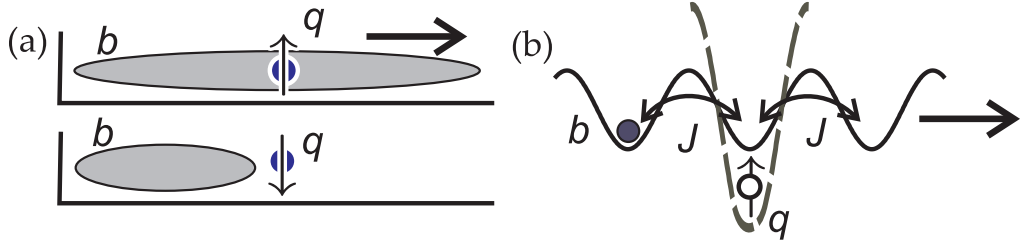}
\caption{(a) A spin 1/2 impurity used as a switch: in one spin
state it is transparent to the probe atoms, but in the other it
acts as a single atom mirror. (b) Implementation of the SAT as a
separately trapped impurity $q$ with probe atoms $b$ in an optical
lattice.} \label{Fig:setup}
\end{center}
\end{figure}

As a second example, we consider a spin-1/2 atomic impurity which
is used to switch the transport of either a 1D Bose-Einstein
Condensate (BEC) or a 1D degenerate Fermi gas initially situated
to one side of the impurity \cite{MicheliETAL:04}. In one spin
state the impurity is transparent to the probe atoms, while in the
other it acts as single atom mirror prohibiting transport.
Observation of the atomic current passing the impurity can then be
used as a quantum non-demolition measurement of its internal
state, which can be seen to encode a qubit,
$|\psi_{q}\rangle=\alpha|\!\!\uparrow\rangle+\beta
|\!\!\downarrow\rangle$. If a macroscopic number of atoms pass the
impurity, then the system will be in a macroscopic superposition,
$|\Psi(t)\rangle
=\alpha|\!\!\uparrow\rangle|\phi_{\uparrow}(t)\rangle+\beta|\!\!\downarrow
\rangle|\phi_{\downarrow}(t)\rangle$, which can form the basis for
a single shot readout of the qubit spin. Here,
$|\phi_{\sigma}(t)\rangle$ denotes the state of the probe atoms
after evolution to time $t$, given that the qubit is in state
$\sigma$ (Fig.~\ref{Fig:setup}a). In view of the analogy between
state amplification via this type of blocking mechanism and
readout with single electron transistors (SET) used in solid state
systems, we refer to this setup as a Single Atom Transistor (SAT).

We consider the implementation of a SAT using cold atoms in 1D
optical lattices: probe atoms in state $\st{b}$ are loaded in the
lattice to the left of a site containing the impurity atom
$\st{q}$, which is trapped by a separate (e.g., spin-dependent)
potential (Fig.~\ref{Fig:setup}b). The passage of $\st{b}$ atoms
past the impurity $q$ is then governed by the spin-dependent
effective collisional interaction
$H_{\mathrm{int}}=\sum_{\sigma}U_{\mathrm{eff,\sigma}}\hat{b}_{0}^{\dag}%
\hat{b}_{0}\hat{q}_{\sigma}^{\dag}\hat{q}_{\sigma}$. By making use
of a quantum interference mechanism, we engineer complete blocking
(effectively $U_{\mathrm{eff}}\rightarrow\infty$) for one spin
state and complete transmission ($U_{\mathrm{eff}}\rightarrow0$)
for the other.

\begin{figure}[ptb]
\begin{center}
\includegraphics[width=13.5cm]{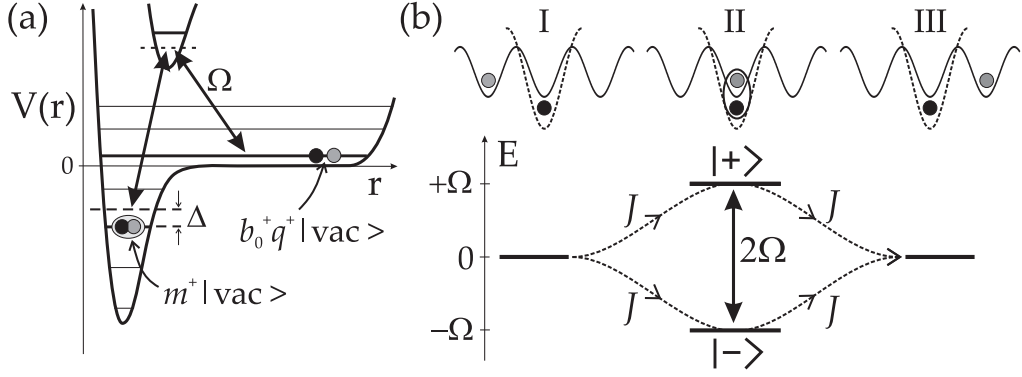}
\caption{(a) The optical Feshbach setup couples the atomic state
$\hat {b}^{\dag}_{0}\hat{q}^{\dag}_{\sigma}|\mathrm{vac}\rangle$
(in a particular motional state quantized by the trap) to a
molecular bound state of the Born-Oppenheimer potential,
$m^{\dag}_{\sigma}|\mathrm{vac}\rangle$, with effective Rabi
frequency $\Omega_{\sigma}$ and detuning $\Delta_{\sigma}$. (b) A
single atom passes the impurity (I$\rightarrow$III) via the two
dressed states (II),
$|+\rangle=\hat{b}^{\dag}_{0}\hat{q}^{\dag}_{\sigma}
|\mathrm{vac}\rangle+m^{\dag}_{\sigma}|\mathrm{vac}\rangle$ and
$|-\rangle
=\hat{b}^{\dag}_{0}\hat{q}^{\dag}_{\sigma}|\mathrm{vac}\rangle-m^{\dag
}_{\sigma}|\mathrm{vac}\rangle$. Quantum interference between the
paths gives rise to an effective tunnelling rate
$J_{\mathrm{eff,\sigma}}$.} \label{Fig:EIT}
\end{center}
\end{figure}

The quantum interference mechanism needed to engineer
$U_{\mathrm{eff}}$ can be produced using an optical or magnetic
Feshbach resonance \cite{Julienne2,Theis,CalarcoMarker}, and we
use the present example to illustrate Hamiltonians for impurity
interactions involving Feshbach resonances and molecular
interactions. For the optical case a Raman laser drives a
transition on the impurity site, $0$, from
the atomic state $\hat{b}_{0}^{\dag}\hat{q}_{\sigma}^{\dag}|\mathrm{vac}%
\rangle$ via an off-resonant excited molecular state to a bound
molecular state back in the lowest electronic manifold
$\hat{m}_{\sigma}^{\dag }|\mathrm{vac}\rangle$
(Fig.~\ref{Fig:EIT}a). We denote the effective two-photon Rabi
frequency and detuning by $\Omega_{\sigma}$ and $\Delta _{\sigma}$
respectively. The Hamiltonian for our system is then given by
$\hat{H}=\hat{H}_{b}+\hat{H}_{0}$, with
\begin{eqnarray}
\hat{H}_{b} &  =-J\sum_{\langle ij\rangle}{\hat{b}}_{i}^{\dag}{\hat{b}}%
_{j}+\frac{1}{2}U_{bb}\sum_{j}{\hat{b}}_{j}^{\dag}{\hat{b}}_{j}\left(
{\hat{b}}_{j}^{\dag}{\hat{b}}_{j}-1\right)  \\
\hat{H}_{0}= &  \sum_{\sigma}\left[  \Omega_{\sigma}\left(
{\hat{m}}_{\sigma
}^{\dag}{\hat{q}}_{\sigma}{\hat{b}}_{0}+\mathrm{h.c}\right)
-\Delta_{\sigma
}{\hat{m}}_{\sigma}^{\dag}{\hat{m}}_{\sigma}\right]  \\
&  +\sum_{\sigma}\left[
U_{qb,\sigma}{\hat{b}}_{0}^{\dag}{\hat{q}}_{\sigma
}^{\dag}{\hat{q}}_{\sigma}{\hat{b}}_{0}+U_{bm,\sigma}{\hat{b}}_{0}^{\dag}%
{\hat{m}}_{\sigma}^{\dag}{\hat{m}}_{\sigma}{\hat{b}}_{0}\right]
,\nonumber \label{HSAT}
\end{eqnarray}
Here $\hat{H}_{\mathrm{b}}$ is a familiar Hubbard Hamiltonian for
atoms in state $\st{b}$; $\hat{H}_{0}$ describes the additional
dynamics due to the impurity on site 0, where atoms in state
$\st{b}$ and $\st{q}$ are converted to a molecular state with
effective Rabi frequency $\Omega_{\sigma}$ and detuning
$\Delta_{\sigma}$, and the last two terms describe background
interactions, $U_{\alpha\beta,\sigma}$ for two particles
$\alpha,\beta\in\{q_{\sigma},b,m\}$, which are typically weak.
This model is valid for
$U_{\alpha\beta},J,\Omega,\Delta\ll\omega_T$ (where $\omega_T$ is
the energy separation between Bloch bands). Because the dynamics
for the two spin channels $q_{\sigma}$ can be treated
independently, in the following we will consider a single spin
channel, and drop the subscript $\sigma$.

To understand the qualitative physics behind the above
Hamiltonian, let us consider the molecular couplings and
associated effective interactions between the $\st{q}$ and
$\st{b}$ atoms for (i) off-resonant ($\Omega\ll\left|
\Delta\right| $) and (ii) resonant ($\Delta=0$) laser driving. In
the first case the effective interaction between $\st{b}$ and
$\st{q}$ atoms is $U_{\mathrm{eff}}=U_{qb}+\Omega ^{2}/\Delta$,
where the second term is an AC Stark shift which plays the role of
the resonant enhancement of the collisional interactions between
$\st{b}$ and $\st{q}$ atoms due to the optical Feshbach resonance.
For resonant driving ($\Delta=0$) the physical mechanism changes.
On the impurity site, laser driving mixes the states
$\hat{b}_{0}^{\dag}\hat{q}^{\dag}|\mathrm{vac} \rangle$ and
$m^{\dag}|\mathrm{vac}\rangle$, forming two dressed states with
energies
$\varepsilon_{\pm}=(U_{qb})/2\pm(U_{qb}^{2}/4+\Omega^{2})^{1/2}$
(Fig.~\ref{Fig:EIT}b, II). Thus we have two interfering quantum
paths via the two dressed states for the transport of $\st{b}$
atoms past the impurity. In the simple case of weak tunneling
$\Omega\gg J$ and $U_{qb}=0$ second order perturbation theory
gives for the effective tunnelling $
J_{\mathrm{eff}}=-\frac{J^{2}}{\varepsilon+\Omega}-\frac{J^{2}}{\varepsilon
-\Omega}\rightarrow0\quad(|\varepsilon|\ll\Omega) $ which shows
destructive quantum interference, analogous to the interference
effect underlying Electromagnetically Induced Transparency (EIT)
\cite{eit}, and is equivalent to having an effective interaction
$U_{\mathrm{eff}}\rightarrow \infty$.

In Ref.~\cite{MicheliETAL:04} the exact dynamics for scattering of
a single $\st{b}$ atom from the impurity is solved exactly,
confirming the above qualitative picture of EIT-type quantum
interference. Furthermore, in this reference a detailed study of
the time-dependent many body dynamics based on the 1D Hamiltonian
(\ref{HSAT}) is presented for interacting many-particle systems
including a 1D Tonks gas.

\section{Conclusions}
\label{Concl}

The cold atom Hubbard toolbox presented in this paper opens a
range of novel and exciting prospects for controlling and
manipulating strongly correlated atomic systems. The resulting
standard and exotic many body Hamiltonians can be used to simulate
otherwise untractable systems in condensed matter physics. The
{\em quantum simulation} of such systems becomes feasible with
cold atoms and provides a valuable tool which can be used in
addition to standard analytical and numerical methods for
analyzing many body systems. In addition we have shown that they
also pose new challenges like e.g.~understanding and utilizing the
unitary dynamics that can be realized in optical lattices.


\end{document}